\documentclass{aastex701}
\usepackage{graphicx} 
\usepackage{subcaption} 
\usepackage{threeparttable}
\usepackage{enumitem}
\usepackage{makecell} 
\usepackage{siunitx} 
\usepackage{amsmath}
\DeclareSIUnit\year{yr}

\DeclareSIUnit \parsec {pc}
\DeclareSIUnit \Jy {Jy}
\DeclareSIUnit \DM {\parsec\per\cubic\cm}

\begin{document}

\title{Back-End System of BURSTT}


\author[0000-0002-8698-7277]{Kai-Yang Lin}
\email[show]{kylin@asiaa.sinica.edu.tw}
\affiliation{Institute of Astronomy and Astrophysics, Academia Sinica, Astronomy-Mathematics Building, No.~1, Sec.~4, Roosevelt Rd., Taipei 106319, Taiwan}
\correspondingauthor{Kai-Yang Lin}

\author[0000-0003-4058-3303]{Chih-Yi Wen}
\affiliation{Institute of Astronomy and Astrophysics, Academia Sinica, Astronomy-Mathematics Building, No.~1, Sec.~4, Roosevelt Rd., Taipei 106319, Taiwan}
\email{wency@asiaa.sinica.edu.tw}

\author[0000-0002-9893-2433]{Homin Jiang}
\affiliation{Institute of Astronomy and Astrophysics, Academia Sinica, Astronomy-Mathematics Building, No.~1, Sec.~4, Roosevelt Rd., Taipei 106319, Taiwan}
\email{homin@asiaa.sinica.edu.tw}

\author[0000-0003-4708-5964]{Andrew Wang}
\affiliation{Institute of Astronomy and Astrophysics, Academia Sinica, Astronomy-Mathematics Building, No.~1, Sec.~4, Roosevelt Rd., Taipei 106319, Taiwan}
\email{andrew@asiaa.sinica.edu.tw}

\author[0000-0003-0340-0651]{Sujin Eie}
\affiliation{Institute of Astronomy and Astrophysics, Academia Sinica, Astronomy-Mathematics Building, No.~1, Sec.~4, Roosevelt Rd., Taipei 106319, Taiwan}
\email{slee@asiaa.sinica.edu.tw}

\author[0000-0002-0060-7975]{Shih-Hao Wang}
\affiliation{Institute of Astronomy and Astrophysics, Academia Sinica, Astronomy-Mathematics Building, No.~1, Sec.~4, Roosevelt Rd., Taipei 106319, Taiwan}
\affiliation{Leung Center for Cosmology and Particle Astrophysics, National Taiwan University, Taipei 106319, Taiwan}
\email{shwang@asiaa.sinica.edu.tw}

\author[0000-0002-6495-8600]{Yao-Huan Tseng}
\affiliation{Institute of Astronomy and Astrophysics, Academia Sinica, Astronomy-Mathematics Building, No.~1, Sec.~4, Roosevelt Rd., Taipei 106319, Taiwan}
\email{yhtseng@asiaa.sinica.edu.tw}

\author[0000-0003-4238-6469]{Hsien-Chun Tseng}
\affiliation{Institute of Astronomy and Astrophysics, Academia Sinica, Astronomy-Mathematics Building, No.~1, Sec.~4, Roosevelt Rd., Taipei 106319, Taiwan}
\email{hctseng@asiaa.sinica.edu.tw}

\author[0000-0001-7888-3470]{Daniel Baker}
\affiliation{Institute of Astronomy and Astrophysics, Academia Sinica, Astronomy-Mathematics Building, No.~1, Sec.~4, Roosevelt Rd., Taipei 106319, Taiwan}
\email{dbaker@asiaa.sinica.edu.tw}

\author[0000-0003-2155-9578]{Ue-Li Pen}
\affiliation{Institute of Astronomy and Astrophysics, Academia Sinica, Astronomy-Mathematics Building, No.~1, Sec.~4, Roosevelt Rd., Taipei 106319, Taiwan}
\affiliation{Canadian Institute for Theoretical Astrophysics, 60 St. George Street, Toronto, ON M5S 3H8, Canada}
\affiliation{Canadian Institute for Advanced Research, MaRS Centre, West Tower, 661 University Avenue, Suite 505}
\affiliation{Dunlap Institute for Astronomy and Astrophysics, University of Toronto, 50 St George Street, Toronto, ON M5S 3H4, Canada}
\affiliation{Perimeter Institute of Theoretical Physics, 31 Caroline Street North, Waterloo, ON N2L 2Y5, Canada}
\email{pen@asiaa.sinica.edu.tw}

\collaboration{all}{BURSTT Collaboration}

\begin{abstract}

The Bustling Universe Radio Survey Telescope in Taiwan (BURSTT) is a new-generation wide-angle radio telescope specifically designed to survey Fast Radio Bursts (FRBs), energetic millisecond-duration pulses of unknown extragalactic origin. 
To realize its scientific potential, which includes detecting approximately 50 FRBs per year and sub-arcsecond localization capability, the system is designed to perform real-time beamforming and pulse search over the \SI{60}{\degree} $\times$ \SI{120}{\degree} field of view.
This paper provides a detailed account of the design, implementation, and performance validation of the BURSTT back-end System. The system employs an efficient multi-stage processing architecture: initial beamforming is executed on the Xilinx ZCU216 RF System-on-Chip (RFSoC) platform; data is then transferred to Intel Xeon servers, where AVX-512 and AMX instruction sets are utilized for the second stage of beamforming and channelization, achieving high computational efficiency to ensure real-time capability. A highly optimized \texttt{bonsai} de-dispersion algorithm performs a real-time pulse search and triggering across 256 beams, which, upon detection, issues commands to the distributed outrigger system to save voltage data for very-long baseline interferometry (VLBI) precise localization. System performance has been validated through beamforming tests using bright radio sources and real-time detection of known pulsars, confirming the high fidelity of the signal processing pipeline.

\end{abstract}



\section{Introduction}

Fast Radio Bursts (FRBs) are energetic, millisecond-duration radio pulses of unknown extragalactic origin, whose physical mechanism and source remain one of the most compelling mysteries in astrophysics. High-sensitivity and high-resolution detection of these transient radio phenomena are crucial for understanding extreme cosmic physics and the distribution of baryonic matter in the Universe.

The Bustling Universe Radio Survey Telescope in Taiwan (BURSTT) \citep{BURSTT2022} is a new-generation radio telescope specifically designed for this purpose. 
It consists of multiple beamforming antenna arrays operating in the 300-800 MHz band, distributed over distances greater than 100 km. 
BURSTT's key advantage lies in its ability to provide an extremely wide instantaneous field of view ($\sim60^{\circ}\times120^{\circ}$, approximately 35 times wider than its predecessor) and sub-arcsecond angular resolution, which is essential for the precise localization of nearby FRB sources that offer the best opportunity for counterpart identification. 
BURSTT is optimized to detect FRBs in the nearby universe (median redshift $z\sim0.04$), with an expected annual detection rate of about 50 events per year, facilitating detailed studies of their properties and origins. The telescope commenced initial FRB search operations in April 2025. 


This paper provides a detailed account of the design, implementation, and performance validation of the BURSTT back-end system. The system employs an efficient multi-stage processing architecture:

\begin{itemize}
    \item \textbf{Initial Processing (Stage 1):} Digital signal processing (F-engine) and the first stage of beamforming are executed on the RF System-on-Chip (RFSoC) platform (using Xilinx ZCU216).
    \item \textbf{Secondary Processing (Stage 2):} Data is then transferred to Intel Xeon servers, where vectorized AVX-512 and AMX instruction sets are utilized for the second stage of beamforming and channelization, achieving over 600~\% of theoretical efficiency to ensure real-time processing capability.
    \item \textbf{Real-time Detection:} A highly optimized bonsai de-dispersion algorithm continuously performs a real-time pulse search and triggering across 256 beams.
    \item \textbf{Localization:} Upon a trigger, the system issues commands to the widely distributed outrigger system to save data, enabling Very-Long Baseline Interferometry (VLBI) for precise localization.
\end{itemize}

The performance of the system has been confirmed through beamforming tests using bright radio sources and successful real-time detection of known pulsars, validating the high fidelity of the signal processing pipeline.


This paper is organized as follows: Section~\ref{sec:MainArray} introduces the top-level data flow architecture of the main array system and the ring buffer design, which is fundamental to high data rate processing.
Section~\ref{sec:beamform_FPGA} details the implementation methods and specifications for the first stage of beamforming performed by the RFSoC-based F-engine on the FPGA.
Section~\ref{sec:beamform_server} discusses the specific methodology and performance evaluation of the efficient second-stage beamforming and channelization implemented on Intel Xeon servers using vector extension instruction sets.
Section~\ref{sec:PulseSearch} describes the bonsai detection pipeline for real-time pulse searching and the multi-beam triggering criteria used for radio frequency interference (RFI) mitigation.
Section~\ref{sec:outrigger} introduces the outrigger system, which includes both domestic and international stations, and its role in FRB localization.
Section~\ref{sec:result} presents the results of system field tests, including beamforming verification using known radio sources and real-time detection performance tests using injected pulses.

In addition to this paper, an overview of the initial operation is reported in \cite{BURSTT_overview} (in preparation). The characterization of the front-end system and the calibration process are summarized in \cite{BURSTT_frontend} (in preparation). The clock synchronization between the stations and the localization verification will be presented in a forthcoming work.


\section{Main array system}
\label{sec:MainArray}

\begin{figure}
    \centering
    \includegraphics[width=0.8\linewidth]{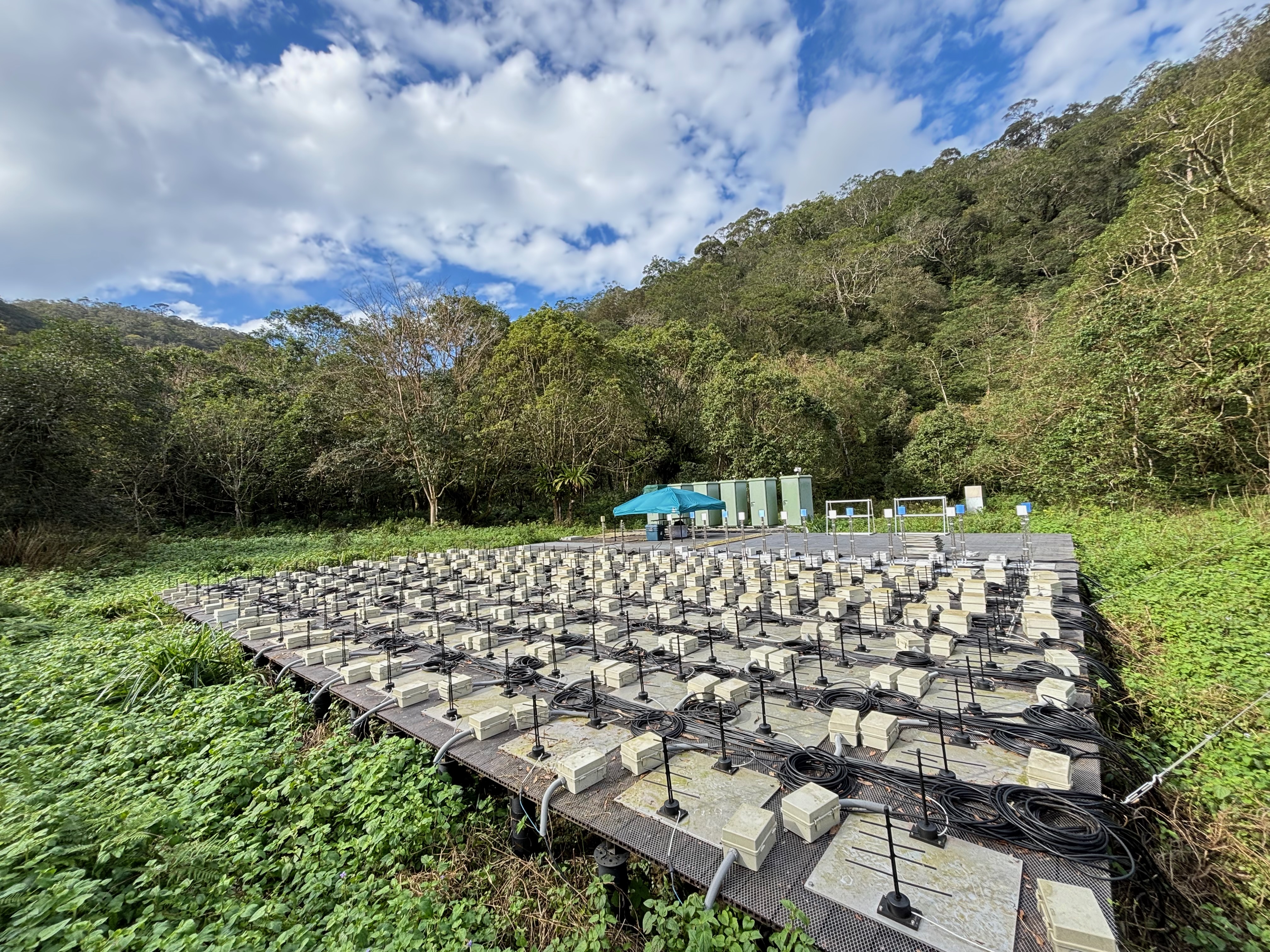}
    \caption{Photo of the current 256-ant main array in Fushan.}
    \label{fig:main_photo}
\end{figure}

The first phase of BURSTT consists of a main array with 256 dipole antennas and several outrigger arrays, each with 64 antennas. Real-time beamforming and pulse detection are carried out with the main array system. 
Upon detection of a dispersed pulse event, a trigger is sent to the main and outrigger arrays to save the baseband data from their ring buffers. 
Figure~\ref{fig:main_photo} shows the main array in Fushan Botanical Garden in the northeastern part of Taiwan. It is a rectangular array with its X and Y axes roughly aligned to the East and North, respectively. In the first phase of operation, the arrays have a few degree misalignment to the Celestial North and East. For the main array, the misalignment is \SI{3.0}{degree}. We plan to reduce the misalignment to the order of \SI{0.1}{degree} with the new arrays in the next phase. The separation of antennas is \SI{1.0}{m} in the X direction and \SI{0.5}{m} in the Y direction. The linearly polarized dipole is aligned to the X direction. Each row (X direction) of sixteen antennas is connected to one Field Programmable Gate Array (FPGA)-based Radio Frequency System on Chip (RFSoC) board (referred to as FPGA boards in the remainder of the text), which performs the digitization, channelization, beamforming (see, e.g. \citealt{Ng2017} in the first dimension (X), and data packetization. 
Data from the FPGA are then sent through a network switch to the beamforming servers that perform beamforming in the second dimension (Y), spectral up-channelization, and integration. 
This second beamforming process occurs independently for each frequency channel. This approach is similar in spirit to the 2D-FFT telescope \citep{Tegmark2009} , although with an implementation that is tailored to our F-engine design. 
After integration, the intensity beams are then streamed to the pulse-search server for real-time pulse detection.

The factorized beamforming process can be expressed by the equation:

\begin{equation}
     B(\theta^X_m, \theta^Y_n,\nu;t) = \sum_j S_{nj}(\theta^Y_n,\nu) \sum_k F_{mk}(\theta^X_m,\nu) V(x_k,y_j,\nu;t) \label{eq:beamforming}   
\end{equation}

\begin{align}
    &F_{mk}(\theta^X_m, \nu) = A_k(\nu) exp(\frac{j2\pi}{\lambda} x_k \sin(\theta^X_m))\,,
    &\sin(\theta^X_m) = \frac{\lambda_0 m}{M \Delta X} + O^X \label{first_beamform} \\
    &S_{nj}(\theta^X_n, \nu) = R_j(\nu) exp(\frac{j2\pi}{\lambda} y_j \sin(\theta^Y_n))\,,
    &\sin(\theta^Y_n) = \frac{\lambda_0 n}{N \Delta Y} + O^Y \label{second_beamform}
\end{align}

$F_{mk}$ and $S_{nj}$ are the first and second beamform matrices, respectively, $V$ is the antenna voltage, $B$ is the beamformed voltage, and $\nu$ stands for the spectral frequency. The indices $j$ and $k$ run from 1 to 16, corresponding to the number antennas in a row and the number of rows, for an array of $16\times16$ antennas. The number of beams to form is an arbitrary parameter limited only by the computing power. In the current phase, BURSTT-256 adopts 16 directions in each of $\theta^X_m$ and $\theta^Y_n$ (i.e. for a total of 256 beams). The indices $m$ and $n$ thus run from 1 to 16. The choice of the beamform matrices were also specified above, where the beam spacing is defined by the reference wavelength $\lambda_0 = c/\mathrm{400~MHz}$ and the spacing $\Delta X = \mathrm{1.0~m}$, $\Delta Y = \mathrm{0.5~m}$. The arbitrary offset terms $O^X$ and $O^Y$ are applied to shift the sky-coverage when needed. They are usaully set to evenly distribute the beams with respect to zenith. A demonstration of the beam distribution can be found in Figure~\ref{fig:256beams}. The factors $A_k(\nu)$ and $R_j(\nu)$ denote the complex weighting factors for each antenna and each row, respectively. These factors act to compensate the instrumental delays, normalize the bandpass function for each antenna and weight each antenna by their sensitivity. The delay and sensitivity weighting can be derived from the eigenvector of the covariance matrix between antennas when the Sun is in the field of view.
In addition, the $A_k(\nu)$ factors, being applied to channelized data with an 18-bit depth in the FPGA, has the potential to mitigate RFI. The RFI mitigation is currently not implemented. Further investigation is deferred to another work.

To elaborate on the beamform process, within each FPGA, a $16\times16$ complex matrix is multiplied to the sixteen antenna voltage streams to produce sixteen X-beam voltage streams. The matrix element can be arbitrarily assigned, and is independent for each spectral channel. Similarly, in the server, another $16\times16$ matrix is applied to the same X-beam from the sixteen FPGAs to further split the X-beam into sixteen Y-beams. The operation is applied to each of the sixteen X-beams to create the final 256 output beams. Since the matrix element is arbitrary, one can perform a discrete Fourier Transformation (DFT) combination of the input streams and retain the full sky information. The intensity beams can be reconstructed from the DFT beams following \citet{Ng2017}, with the caveat that the spectral sensitivity variation depends strongly on where the source is within the beam. For simplicity, we have chosen to compensate the geometric delay of every antenna to a specific direction on the sky for a selected beam. Thus, the sensitivity varies smoothly across the spectral range and does not strongly depend on the source direction within the beam. However, as the system covers a wide frequency band (\SIrange[]{400}{800}{MHz}), the beam size at the higher frequency end is naturally smaller than at the lower frequency end. Therefore, a source at the outskirt of the beam will have less sensitivity at higher frequency compared to when the source is at the beam center. After detection, the source spectral property can be reconstructed using the baseband data in an offline beamform process. On the other hand, as the effective collecting area of a dipole antenna is proportional to the wavelength squared, the higher frequency band is much less sensitive compared to the lower frequency band. As a consequence, most of the detection will be obtained from the lower frequency band. In our future expansion, we plan to reduce the frequency coverage in favor of adding more dipoles to optimize the detection sensitivity.

\begin{figure}
    \centering
    \includegraphics[width=0.8\linewidth]{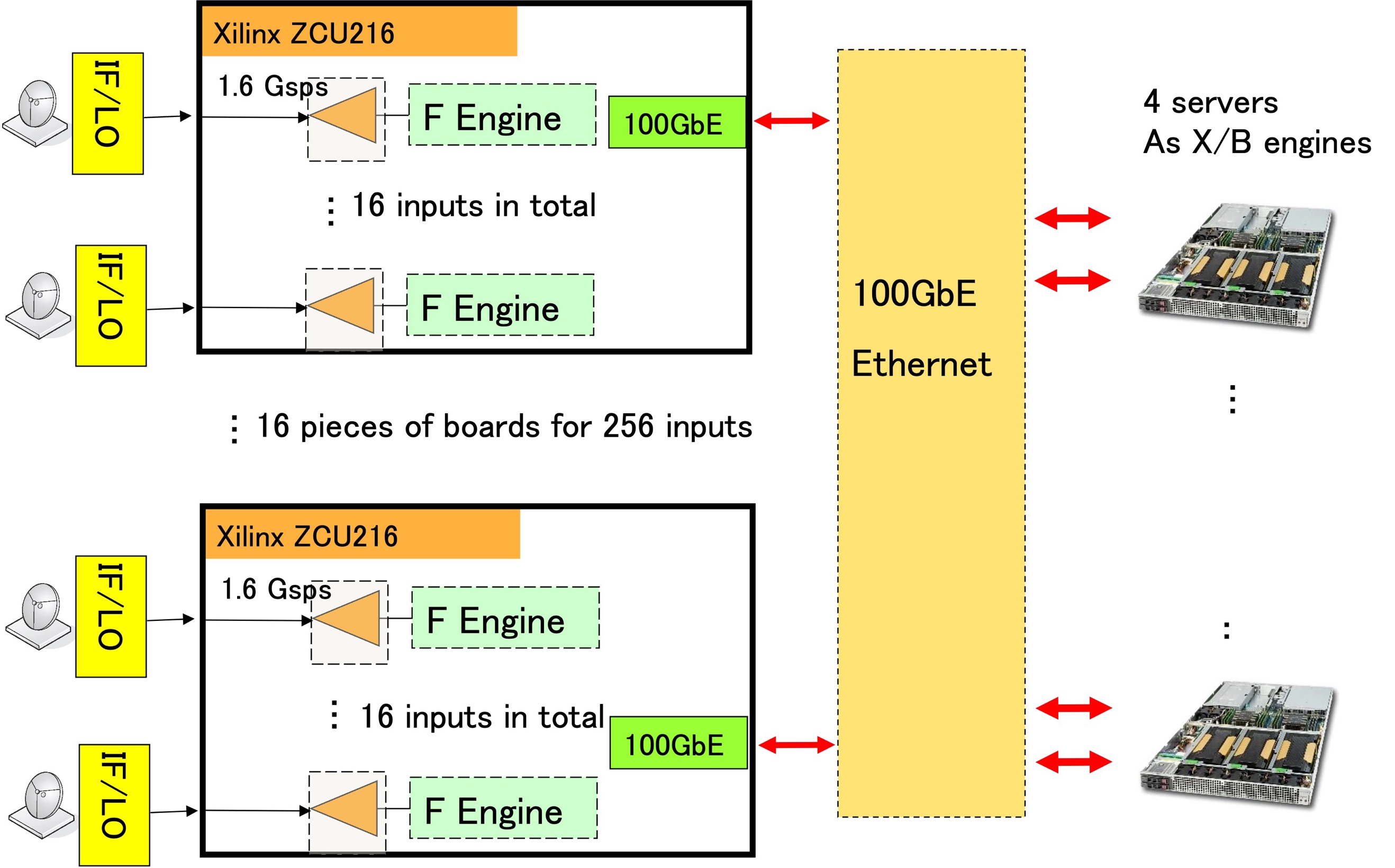}
    \caption{Backend system of BURSTT phase I.}
    \label{fig:backend_block_diagram}
\end{figure}

The main array system contains sixteen FPGAs, four beamforming servers, and one pulse-search server. Data from the FPGA is packaged as user datagram protocol (UDP) packets. 
Each beamforming server contains two non-uniform memory access (NUMA) nodes. Each node consists of a CPU (Intel Xeon 8458p), 1~TB of RAM, and a dual-port 100GbE network interface card (Mellanox Technologies MT28908 Family, ConnectX-6) and will receive 50~MHz of data (i.e. 1/8 of the total bandwidth) from all 16 FPGAs. Totally, there are sixteen 100GbE receiving ports on the beamforming servers, and a 32-port 100GbE network switch is used to distribute the packets. Figure~\ref{fig:backend_block_diagram} shows a schematic diagram of the main array backend system architecture. 

The baseband voltage data and processed intensity data will be saved as ring buffers. The raw intensity is further processed and streamed to the pulse-search server. The pulse-search server receives intensity data from four beamforming servers (eight nodes) and detects the presence of FRB signals. 
Upon detection, a trigger command will be sent back to the beamforming servers and outrigger stations to save the corresponding data segments for further confirmation and VLBI localization.  
 
\begin{figure}
\begin{center}
\resizebox{150mm}{!}{\includegraphics{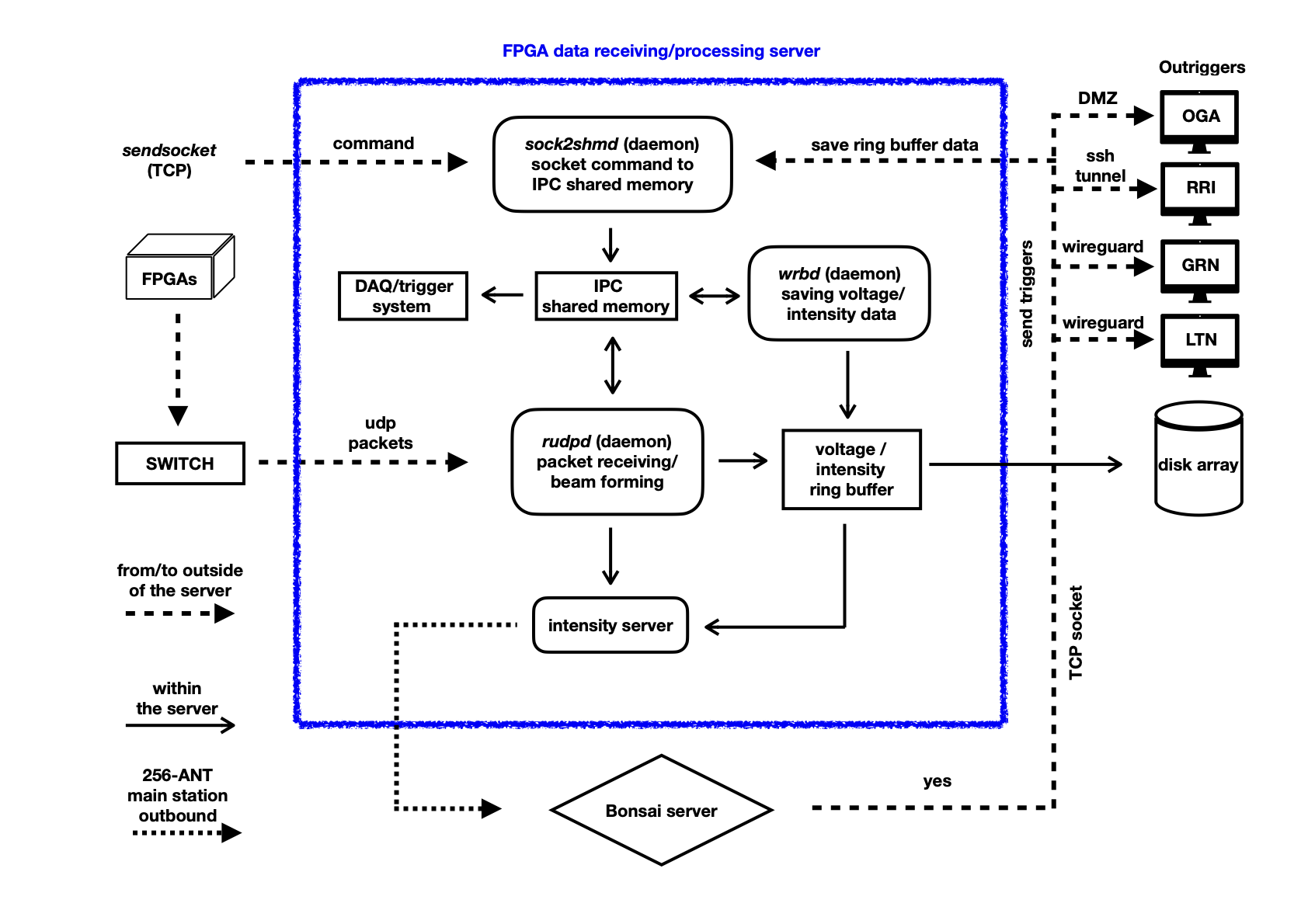}}
\end{center}
\caption{Flow chart of BURSTT backend system.}
\label{Fig:FlowChart}
\end{figure}

Figure~\ref{Fig:FlowChart} shows the system flow chart of the packet receiving/processing system. The design concepts are as follows. A small memory block (called the shared memory) is allocated for inter-process communication (IPC) to maintain the incoming commands through network and share to all processes within the server. This shared memory also maintains  the error codes,  data rate, incoming command, and the system information of that server. All of this information is written to log files every 300 seconds for future reference. 
The packet receiving software system contains the following 4 items, either a software process ({\it sendsocket}) or a daemon ({\it sock2shmd, rudpd, wrbd}) that is run as a background process.
\begin{enumerate}
\item {\it sendsocket} sends start/stop/save commands through network socket with transmission control protocol (TCP) to designated servers to execute corresponding tasks.
\item the socket-to-shared-memory daemon ({\it sock2shmd}) listens to the incoming TCP commands and updates the shared memory. 
\item the read-UDP-packet daemon ({\it rudpd}) starts/stops receiving packets when the corresponding commands appear in the shared memory.
\item the write-ring-buffer daemon ({\it wrbd}) monitors the command in the IPC shared memory for any incoming saving command. Once it receives a save command with estimated dispersion measure (DM) and arrival time from the pulse-search server, the daemon will save the corresponding data segments in the ring buffers. 
\end{enumerate}

In order to improve the memory-access performance, our ring buffer page size is configured to be hugepages with a 1~GB page size to reduce the overhead.  The current ring buffer size is equivalent to 25 seconds in time, which is capable of detecting FRBs with the DM up to $\sim$\SI{1000}{\DM}.

\section{RFSoC beamforming}
\label{sec:beamform_FPGA}

An RFSoC (Radio Frequency System on Chip) is a type of chip that combines digital signal processing and analog-to-digital conversion in one place. It also has digital down-converting and signal attenuator features. RFSoC has made radio systems simpler. It has changed radio astronomy by offering a powerful, low-energy, and small way to handle radio signals. Because of these benefits, it is becoming popular in radio communication systems and radio telescopes. For instance, \citet{Ergesh2022} and \citet{Pei2022} used the 8-channel RFSoC ZCU111 board by Xilinx for their system design.

Reducing the cost per RF channel is very advantageous for low-frequency radio arrays because they usually have hundreds or thousands of antennas. 
The RFSoC product from Xilinx, model XCZU49DR, which has 16 channels of analog-to-digital converter (ADC) inputs and a lot of resources, was quickly identified as a suitable processor. 
Nygaard \citep{Nygaard2021} used Xilinx ZCU216, which is a development kit of XCZU49DR, to develop ALPACA for the Green Bank Telescope and shared their libraries with the Collaboration for Astronomy Signal Processing and Electronics Research (CASPER; \citealt{Hickish2016}) for public use. 
We are designing our model based on the ZCU216 hardware and using CASPER’s RFSoC library to meet our specifications, shown in Table~\ref{tab:F_engine_spec}.

\begin{table}[h]
\centering
\begin{tabular}{lcc}
\hline
\textbf{}  \\
\hline
No of Antenna Input     & 16 \\
ADC bit width    & 14 bits \\
ADC Clock Rate        & \SI{1.6}{\GHz} \\
Digital bandwidth       & \SIrange{0}{800}{\MHz} \\
F-engine clock rate	   & \SI{400}{\MHz}   \\
No. of Spectral Channels	& 2048   \\
Bit length in FFT	& 18 bits   \\   
Re-Quantization after FFT 	& 4 bits + 4 bits   \\
No. of Output Channels & 1024 \\ 
Data transfer rate per 100G cable	& \SI{51.2}{\giga\bit\per\second}
   \\
Total no. of 100G cables	& 1   \\
Synchronization clock rate	& 1 PPS/10MHz from GPS    \\

\hline
\end{tabular}
\caption{Specifications of BURSTT's F-engine unit.}
\label{tab:F_engine_spec}
\end{table}

\begin{figure}
    \centering
    \includegraphics[width=0.8\linewidth]{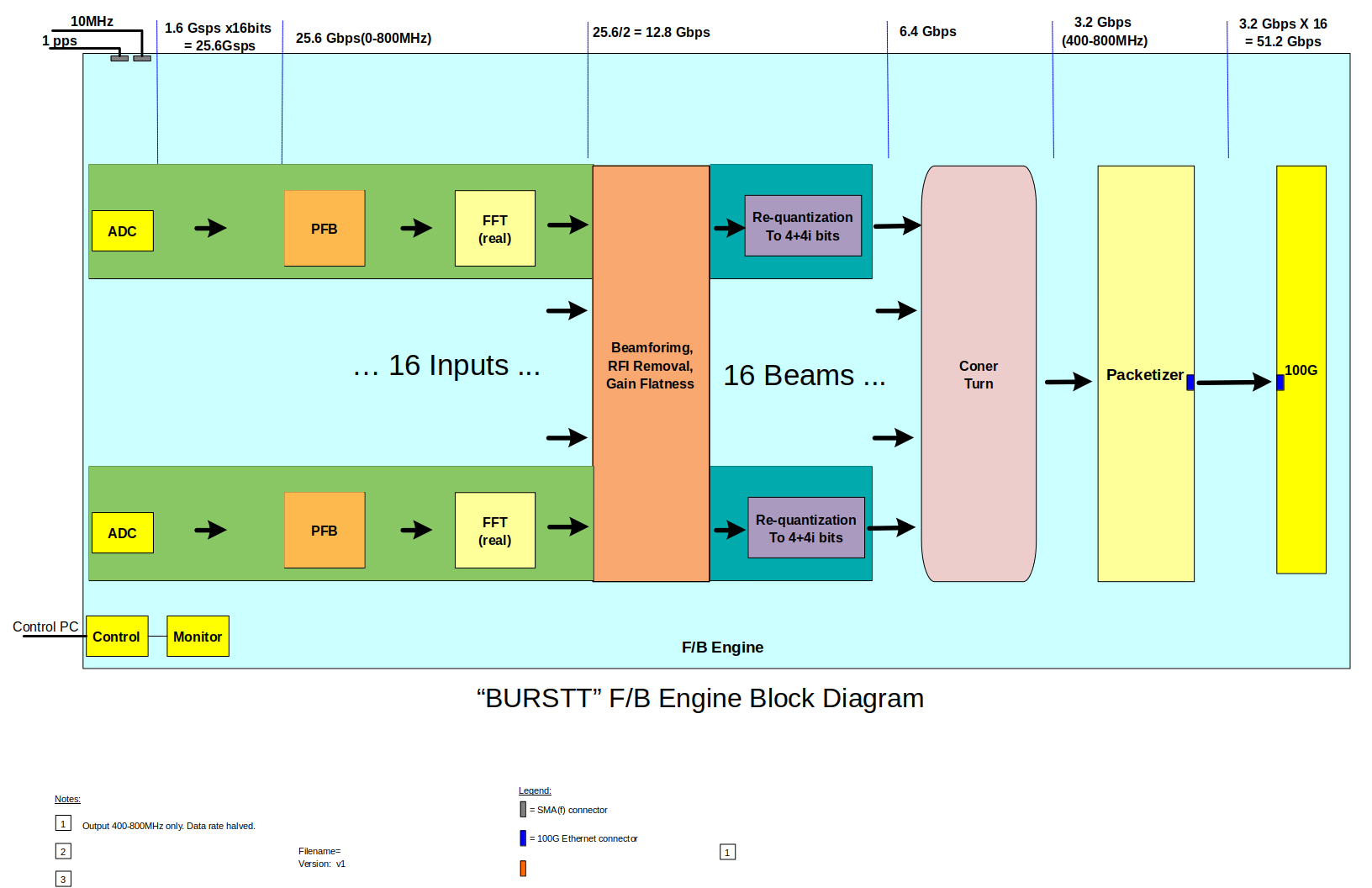}
    \caption{Block diagram of BURSTT’s F-engine with \SI{1600}{\MHz} sampling rate and \SI{800}{\MHz} bandwidth. In the end, data at \SIrange{0}{400}{\MHz} are discarded, and the data rate reduces to \SI{51.2}{\giga\bit\per\second}.}
    \label{fig:F_engine}
\end{figure}

BURSTT deploys 256 antennas, requiring 16 ZCU216 units to process their signals. A 100~Gbps Ethernet switch routes the data packets to the appropriate computer server for cross-correlation and beamforming (Fig.~\ref{fig:backend_block_diagram}). The astronomical signal from the antenna goes to the ADC module in the RFSoC, sampling at 1600~MHz. 
The data is multiplexed by four and sent to the FPGA, which runs at 400~MHz. Using the CASPER GUI DSP library, the data is filtered by a poly-phase filter bank (PFB) and changed from time to frequency domain using fast Fourier transform (FFT) blocks. To facilitate cross-correlation in the beamforming server, a matrix transpose, or corner turn, is performed after FFT (Fig.~\ref{fig:F_engine}). To fit the 100~Gbps Ethernet format, a custom packetizer is made in Simulink. To minimize the packetizing overhead, 8~kB of data is stored in each UDP packet. Fig.~\ref{fig:F_engine} shows the model design.

Beamforming, by combining signals from multiple antenna elements, creates a directional beam focused on a specific region of the sky, thereby reducing the impact of unwanted background noise and interference. This process facilitates the detection of weak astronomical signals \citep{Warnick2018}. Beamforming enhances the spatial resolution of radio telescopes, which is crucial for the study of celestial objects. By combining signals from multiple antennas with known relative positions, beamforming can simulate a virtual telescope with an effective aperture, enabling fine angular resolution. \citet{Watkins2002} demonstrated real-time beamformer QR decomposition on an FPGA platform \citep{Dick2007}.

 Radio scientists implement beamforming using a PC equipped with a graphics processing unit (GPU), as seen in the ALPACA \citep{Nygaard2021} and CHIME telescope \citep{Denman2020}. However, the BURSTT beamforming computation, which uses only matrix multiplication, is modest compared to the full cross-correlation needed for CHIME. The computation can be easily shared by the FPGAs and the CPUs. Therefore, the power of passing data into GPUs for a large computation would be wasted in BURSTT. We have developed an FFT digital backend system for the BURSTT based on the RFSoC platform. This design is based on the open-source radio astronomical library developed by the Collaboration for Astronomy Signal Processing and Electronics Research (CASPER) \citep{Hickish2016}. Our beamforming model with 1K, as illustrated in Fig.~\ref{fig:F_engine}, processes the 16 generated beams through a corner turn block in 100~Gbps Ethernet format.	
 
The weighting factors shown in Equation~\ref{eq:beamforming} are derived offline using calibration observations of the Sun (\citealp{BURSTT_frontend}, in preparation). The complex weighting factor takes into account the bandpass equalization, instrumental delay compensation, and the beam-steering delay compensation. 
Optionally, a spatial filtering radio frequency interference (RFI) mitigation matrix can also be included. 
We then consolidated the beamformer weighting factor, the RFI eigenvalues, and the spectrum flatten gains into a single variable. This newly combined variable was transferred to an FPGA in an 8-bit complex format and multiplied by our real-time spectral data. 
To form the 16 beams, a 16$\times$16, i.e., 256, 1 K blocked random access memory (BRAM) FPGA fabric is required.

Given that the FFT output is 18 bits, we utilize a 16-bit weighting factor for beamforming. However, the 16-bit weighting factor nearly exhausted the available Block RAM (BRAM). By reducing the weighting to 8 bits, we utilized only half of the available BRAM, which sufficiently reduced the fabric resource usage to accommodate the ZCU216.

After the beamforming process, the baseband voltage is truncated to 4-bit real and 4-bit imaginary parts. To optimize the limited bit depth, bandpass equalization is a critical step in the beamforming matrix. On the other hand, since we express the final beamforming matrix in a signed 8-bit integer, spectral channels with a smaller gain adjustment (i.e. channels that are originally stronger) will have fewer effective bits to store the matrix parameters. This problem will be more severe when the bandpass has a large gain slope that needs to be compensated. In our current system, there is approximately a 10~dB slope across the band, and the beamforming matrix is expressed by about 6 bits instead of 8 bits.

It is worth noting that there are 1024 spectral channels, with 8~bits per channel, and sixteen output beams, a full spectral frame output takes 16384~bytes. Since this size exceeds the Maximum Transmission Unit (MTU) of about 9000, we use a minimum of 2 packets to transmit each spectral frame. However, in the case of the main array, the loading of server beamforming is shared by eight processing nodes. Each node will receive and beamform data from all FPGA boards in $1/8$ of the bandwidth. Therefore, the spectral frame is split into eight UDP packets, where each spectral subband is directed to one of the nodes through the switch. On the other hand, the subband takes up only 2048~bytes, which is smaller than the MTU and less efficient. We thus accumulate four spectral frames in each UDP packet to optimize the transmission efficiency.

\subsection{Synchronization}\label{sec:synchronization}

Data synchronization between FPGA boards is maintained by a master clock at each site. A \SI{10}{MHz} reference signal and a one-pulse-per-second (1PPS) signal are amplified and distributed to the FPGA boards. Each FPGA board, running at a clock rate of \SI{400}{MHz}, is phase-locked to the external \SI{10}{MHz} reference. The 1PPS is used to synchronize data processing. Upon reception of a RESET command, the data buffer and processing in the FPGA is halted when the next 1PPS arrives and the system is ARMED. The system is TRIGGERED and processing started when the next 1PPS arrives. A spectral frame is generated every 1024 clock cycles (at \SI{400}{MHz}). A counter stores the number of frames produced since the RESET. The RESET time and the frame counter are stored in the UDP packet as a timestamp of the transmitted data. In the server ring buffer, we use the frame counter to align the data received from different FPGA boards. For diagnostic purposes, the validity of the data packets is stored separately in a bitmask within the ring buffer, one bit for each packet. The bitmask is not taken into account in real-time processing. However, there is practically no data loss in the main array system, with a data loss rate less than $2.5e-8$. The payload of each UDP packet is 8256~bytes, where the first 64~bytes contain the timestamps and frequency information. The remaining 8192~bytes are used for the baseband spectra.

\section{Server beamforming} \label{sec:beamform_server}

We perform first-stage beamforming on RFSoCs, followed by second-stage beamforming on Intel Xeon servers. The most efficient way to execute beamforming on CPUs is by leveraging advanced vector or matrix extensions --- specifically, AVX-512 on 3rd Gen Xeons and AMX on 4th Gen Xeons.

On the 3rd Gen Xeon server, we utilize the \texttt{vpdpbusd} instruction from the AVX512-VNNI extension. This instruction operates on two AVX-512 registers containing sixty-four 8-bit integers each and produces one AVX-512 register with sixteen 32-bit integers. It effectively performs 64 multiplications and 48 additions in a single operation. With proper optimization, this instruction can significantly accelerate the matrix multiplications used in beamforming computations.

On the 4th Gen Xeon server, we can further improve performance using the \texttt{tdpbf16ps} instruction from the AMX-BF16 extension. This instruction takes two two-dimensional registers containing 32×16 bf16 elements as inputs and produces a 16×16 fp32 result matrix. By applying two \texttt{tdpbf16ps} instructions -- one for the real part and one for the imaginary part -- we can efficiently compute the multiplication of two 16×16 complex matrices.

The up-channelization process also involves matrix multiplications. It is performed by multiplying a discrete Fourier transform (DFT) matrix with 16-point time-series data from a single channel to produce 16 frequency channels.

In this section we present the performance evaluation of beamforming and up-channelization on the Intel Xeon Platinum 8458P processor. According to Intel's APP Metrics, the peak theoretical performance of this processor is 2252.8 GFLOPS. It features 44 cores, 88 threads, a base clock of 2.7\,GHz, and a turbo frequency up to 3.8\,GHz.\footnote{\url{https://www.intel.com/content/www/us/en/content-details/840270/app-metrics-for-intel-microprocessors-intel-xeon-processor.html}}

Beamforming involves complex $16 \times 16$ matrix multiplications. Each output element requires 126 floating point operations (FLOPs), including complex multiplications and additions. With 256 output elements per matrix, each multiplication requires approximately 32,256 FLOPs.

Incoming data from FPGAs consist of 128 channels sampled at 400,000 samples/s, totaling 51.2 million samples per second. Each sample includes data from 256 antennas, with a size of 256 bytes. This results in a beamforming computational load of:
\[
51.2 \times 10^6 \times 32{,}256 \approx 1651.5 \;\text{GFLOPs/s}.
\]

Compared to the Xeon's theoretical 2252.8 GFLOPS, this accounts for 73.3\% of total capacity, or the equivalent of 32.2 out of 44 available cores.

If up-channelization is included (via a $16 \times 16$ DFT), the computation roughly doubles (3303 GFLOPS and 64.5 cores required)
which exceeds the available 44 cores. To improve the performance of the beamforming pipeline and fit the available 44 cores we provide two implementations of the code.

The beamforming pipeline includes:
\begin{itemize}
  \item Converting and reordering input in 4+4-bit complex format
  \item Second beamforming
  \item Optional up-channelization
  \item Optional single-beam data output
  \item Intensity integration ($T=400$): the intensity of baseband data is integrated and calculated over 400 frames. The intensity data is streamed to pulse-search server for real-time pulse search (see Sec.~\ref{sec:preproc}). 
\end{itemize}

There are two implementations of the code:
\begin{itemize}
  \item \textbf{AVX512 version}, using \texttt{\_mm512\_dpbusd\_epi32}
  \item \textbf{AMX version}, using \texttt{\_tile\_dpbf16ps}
\end{itemize}

Only the multiplication and addition operations for the second and third steps are counted in the FLOPs estimation, so actual performance may be slightly higher due to overhead not accounted for.

Currently, we use two AMX instructions to perform matrix multiplication on two complex 16×16 matrices (bf16), producing two 16×16 real matrices (fp32): one for the real part and one for the imaginary part. To achieve this, we prepare two auxiliary matrices: one with swapped real and imaginary parts, and another with the imaginary part negated. If future Xeon CPUs provide AMX-COMPLEX instructions, these preparation steps will no longer be necessary.

\begin{table}[h]
\centering
\begin{tabular}{lcc}
\hline
\textbf{Implementation} & \textbf{Cores Required} & \textbf{Efficiency vs Theoretical} \\
\hline
AVX512 (beamforming only)     & $\sim$10.2 & 316\% \\
AVX512 + up-channelization    & $\sim$22.8 & 283\% \\
AMX (beamforming only)        & $\sim$5.36 & 601\% \\
AMX + up-channelization       & $\sim$9.80 & 658\% \\
\hline
\end{tabular}
\caption{Measured performance of AVX512 and AMX implementations on Xeon 8458P. Compared to the Xeon's theoretical 2252.8 GFLOPS}
\label{tab:AVX_AMX}
\end{table}

The AMX-based implementation achieves over 600~\% efficiency relative to the theoretical baseline and supports real-time processing even with up-channelization, as summarized in Table~\ref{tab:AVX_AMX}. The AVX512 version also meets real-time requirements and offers good efficiency, though with higher core usage.
These results demonstrate that optimized Xeon implementations can support real-time server-side beamforming and channelization with ample performance margin when using advanced instruction sets like AMX.

\section{Pulse search}
\label{sec:PulseSearch}

\subsection{Preprocessing of Intensity Data Stream}
\label{sec:preproc}

The beamformed intensity is integrated over time to a resolution of $\rm 1.024~ms$. To enable real-time searches up to $\rm DM \gtrsim \SI{1000}{\DM}$, the data are re-channelized from 1024 to 16384 channels using an FFT-based channelization technique, as explained in Section~\ref{sec:beamform_server}. The resulting intensity data are then normalized by the time average of blocks with length $0.524288~\rm s$. To suppress common instrumental RFI appearing across multiple beams, an incoherent beam --- constructed by averaging the intensity of all beams --- is subtracted from each beam. For initial RFI mitigation, a high-pass filter was applied by subtracting a rolling mean with a 31 ms window from each beam’s intensity. The filter size is adjustable; we selected 31 ms to avoid attenuating pulses from radio pulsars while still effectively suppressing noise with durations longer than a few tens of milliseconds. The RFI-mitigated data are then standardized to have unit variance. Finally, sigma clipping with signal-to-noise ratio SNR = 4 is applied to the channelized and normalized intensity to remove strong outliers, including intermittent narrowband RFI spikes, while ensuring that the DM-domain pulse search remains effective for detecting dispersed pulses with significant SNR.

To quantify the level of impulsive time-domain RFI remaining after mitigation, we examined the statistical distribution of the 8-bit intensity samples. In the absence of RFI, the noise is expected to follow a Gaussian distribution, while impulsive interference produces excess non-Gaussian tails and potential digitizer saturation. Across all beams, we measure a fraction of samples exceeding $\pm4\sigma$ of $f_{\rm tail} \approx (1.0\text{--}1.7)\times10^{-4}$, compared to $6.3 \times 10^{-5}$ expected for purely Gaussian noise. The excess therefore corresponds to a non-thermal occupancy of approximately $1 \times 10^{-4}$ of all time samples. We find no evidence of digitizer saturation, with $f_{\rm sat} < 10^{-6}$.
These results indicate that impulsive time-domain RFI affects only $\sim 0.01 \%$ of the data, substantially smaller than the fraction of masked digital TV channels ($\sim 13 \%$ of the search bandwidth), and therefore does not limit burst detectability or bias measured burst properties.

\begin{figure}
    \centering 
    \includegraphics[width=0.8\textwidth]{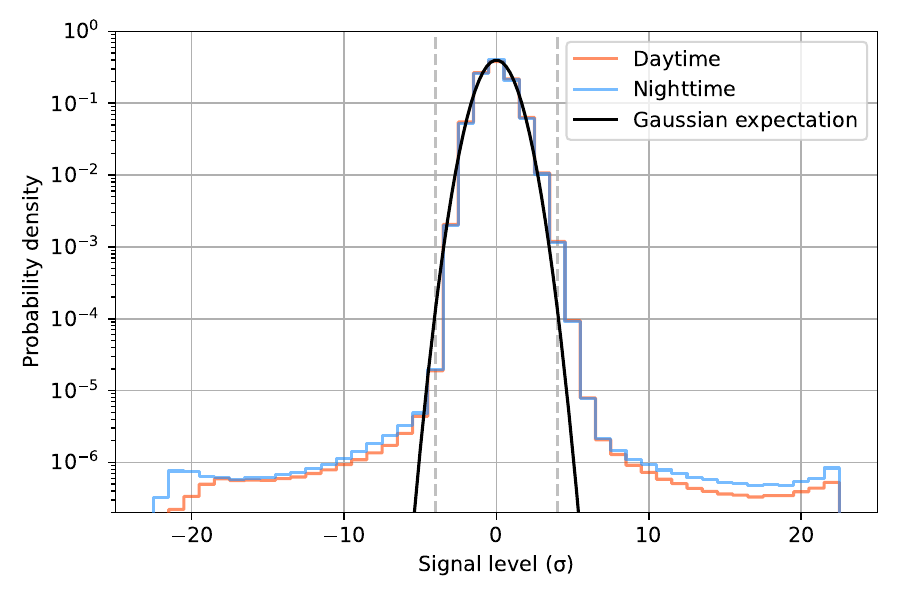}
	\caption{
        Distribution of preprocessed intensity samples before clipping. The black curve shows the expected Gaussian distribution. The small excess in the $\pm 4 \sigma$ tails corresponds to a time-domain RFI occupancy of $\sim 10^{-4}$ of samples.
	}
\label{fig:histogram}
\end{figure}

The pre-processed intensity ring buffer is then forwarded to the real-time search pipeline. We use a highly optimized tree dedispersion
algorithm, \texttt{bonsai} (see \citealt{CHIME2018},
\citealt{Merryfield2023} for details). The system utilizes two
non-uniform memory access (NUMA) nodes equipped with Intel Xeon
Platinum 8593Q CPUs to process BURSTT’s 256 beams. Each beam is
assigned to a dedicated \texttt{bonsai} process, with groups of three
beams processes sequentially within individual CPU cores, enabling
parallelized \texttt{bonsai} executions. In addition, an incoherent
beam is stored in the ring buffer and forwarded to \texttt{bonsai} for
use in the subsequent RFI classification.  

\subsection{\texttt{bonsai} Configuration for Pulse Search}
\label{sec:bonsai_conf}

\begin{figure*}
    \centering
    \includegraphics[width=1.0\textwidth]{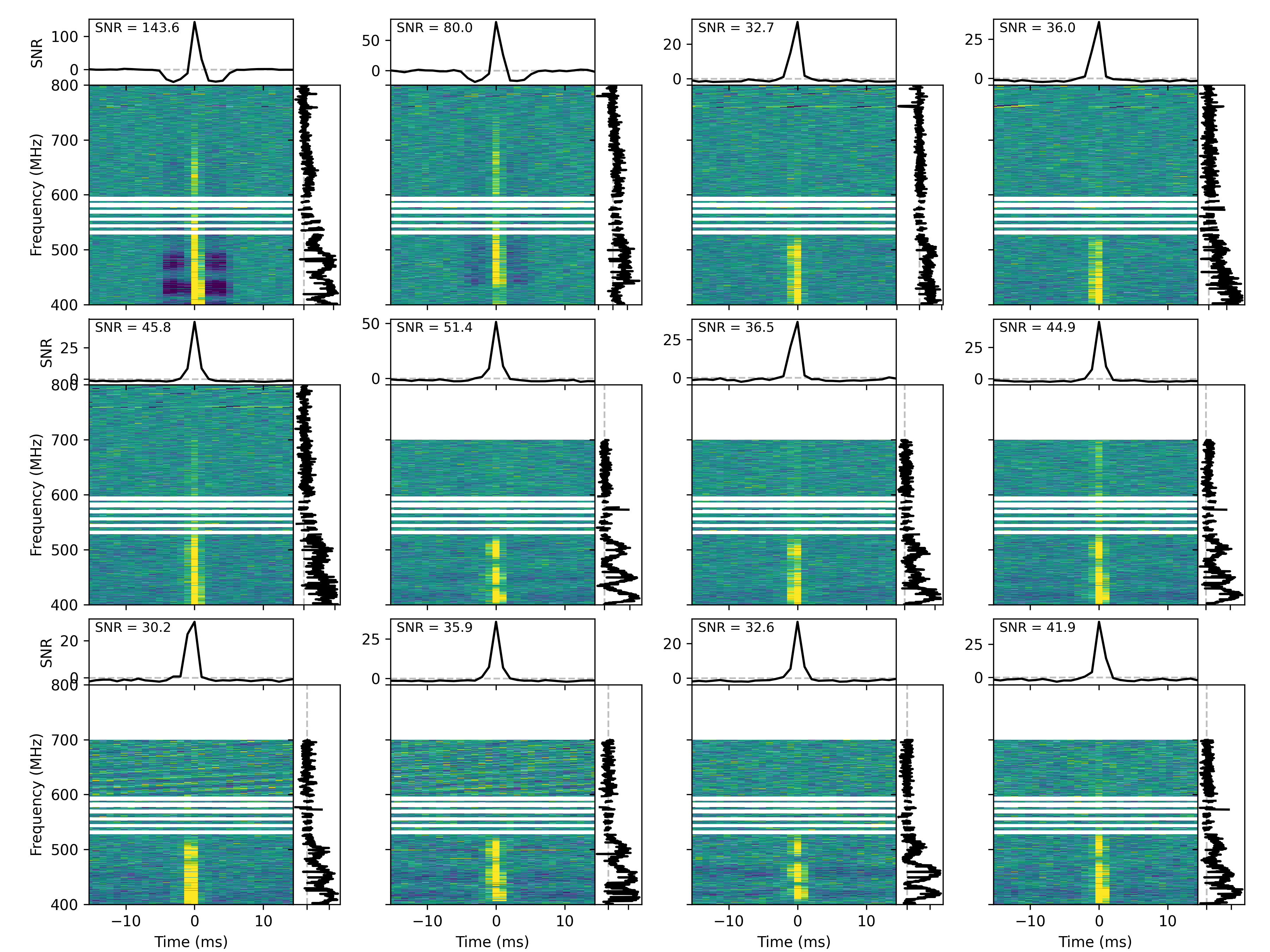}
    \caption{
    Bright giant pulses from the Crab pulsar detected by the BURSTT real-time pipeline. Each main panel shows a dedispersed dynamic spectrum spanning 400--800 MHz. The frequency-averaged profile (400--700 MHz) is shown above, and the time-averaged spectrum is shown to the right for a 3 ms window around the pulse peak. The intensity data plotted here have a time resolution of $\rm \SI{1.024}{\milli\second}$, with frequency binning of 16 across 16,384 channels covering the \SI{400}{\MHz} bandwidth. A rolling mean with a \SI{31}{\milli\second} window was subtracted from the intensity data. For the first two panels, a \SI{9}{\milli\second} high-pass filter was applied instead, during the early commissioning period. All pulses were detected and dedispersed at $\rm DM = \SI{56.76}{\DM}$. The horizontal blank stripes indicate the masked frequency channels due to RFI.
    }
    \label{fig:CrabGRP}
\end{figure*}

Each \texttt{bonsai} process receives the ring buffer and reads out the intensity from one beam from \SI{400}{\MHz} to \SI{700}{\MHz} with a time resolution of \SI{1.024}{\milli\second} and a frequency resolution of \SI{24.4140625}{\kHz}. In total, 12288 out of 16384 frequency bins are employed in the pulse search process. In our pulse search pipeline, the frequency range of \SIrange{700}{800}{\MHz} is excluded because receiving dipole antennas intrinsically exhibit a reduction in effective area with increasing frequency, leading to degraded sensitivity at higher radio frequencies. 
In addition, five digital TV bands between \SIrange{520}{600}{\MHz} are masked, each with a bandwidth of $\sim$\SI{8}{\MHz}, as also shown in Figure~\ref{fig:CrabGRP} and Figure~\ref{fig:B0329}. 
In total, \SI{40}{\MHz} out of \SI{300}{\MHz} bandwidth ($\sim13\%$) is masked.
To enable real-time processing of all 256 beams on a single \texttt{bonsai}-dedicated pulse-search server, time samples are coarse-grained by a factor of 8 to regulate the disk output rate. 

For the time-domain search, each single-tree processes a data block of \SI{0.524288}{\second} and performs an incremental search across the data stream. With the \SIrange{400}{700}{\MHz} search band at the main station, pulses with DMs up to \SI{1000}{\DM} and widths up to \SI{3.072}{\milli\second} are searched with a DM step of \SI{0.058638}{\DM}. A single-beam level (level 1, L1) trigger is raised when an event exhibits a SNR over $8.0$ ($\rm SNR_{th} = 8.0$) and DM over \SI{16}{\DM} at any of the 256 beams. Events meeting these criteria are promoted to the multi-beam trigger stage for further verification. The searched parameter space of DM and arrival time at 400~MHz ($t_0$) for the triggered beam is coarse-grained and saved. The maximum SNR and DM of all the rest of beams, including the incoherent one, are also recorded for trigger processing.


\subsection{Multi-Beam Trigger Criteria and RFI Rejection}
\label{sec:TrigCriteria}


When an L1 trigger is raised by any of the 256 beams, the following selection criteria must be satisfied to issue a multi-beam (level 2, L2) trigger, which is designed to reject RFI:

\begin{enumerate}[label=(\roman*),ref=\roman*]
    \item The L1 trigger has the highest SNR among all beams within the same DM range. Since each beam is searched independently, this criterion filters out detections from all but the brightest beam. Real astrophysical signals tend to be detected in a single beam but can appear in spatially adjacent beams or sidelobes. Terrestrial RFIs mostly appear in multiple beams simultaneously.
    \item The SNR of the incoherent beam $\rm{SNR_{incoh}}$ is lower than a given threshold $\rm{SNR_{th}}$: for a similar reason as above, terrestrial RFIs tend to have significant $\mathrm{SNR_{incoh}}$ values. \label{L2crit:2}
    \item Within the previous five blocks, including the latest block under analysis (corresponding to \SI{2.62}{\second}), the SNR values in the low-DM region ($\mathrm{DM} < 10$) remain below $\mathrm{SNR_{th}}$. \label{L2crit:3}
\end{enumerate}

The events passing L2 are considered as pulse candidates, and the preprocessed intensities of both the event beam and the incoherent beam, along with the coarse-grained 2D array of (DM, $t_0$) and pulse search parameters, are recorded for further investigations. 
On the other hand, events that pass the single-beam level criteria (L1) but fail the multi-beam level criteria (L2, sub-criteria~\ref{L2crit:2} and \ref{L2crit:3}) are classified as RFI. For strong RFI events with $\mathrm{SNR} > 10$, their intensities and corresponding search results are also recorded to monitor ambient noise and refine the trigger criteria.
A web-based monitoring was developed to inspect individual events and their distribution in near real time.

\subsection{Event Identification with Known Pulsars}
\label{sec:EventID}


For an event that passes the L2 trigger, its best-fit DM and the beam position in equatorial coordinates (RA, DEC) corresponding to the maximum SNR are cross-checked against the ATNF Pulsar Catalogue \footnote{\url{https://www.atnf.csiro.au/research/pulsar/psrcat/}} \citep{ATNF_catalogue}.
An event is flagged as a pulsar event if both its DM and coordinates fall within the specified tolerances, $\pm\SI{2}{\DM}$ in DM and within the beam region expanded by an additional $\ang{0.5}$.
The system reaction to each pulsar can be individually defined, e.g., whether to reject or issue triggers to all stations to save the baseband data, or with a different SNR threshold. Currently, BURSTT records the baseband data for all incoming pulsar events except those from PSR B0329+54 which causes frequent trigger. See Sec.~\ref{sec:pulsar} for details.
%
If an event is not rejected at this stage, the trigger signal is distributed to all BURSTT stations, including the (DM, $t_0$) information for saving the baseband data.

\section{Outrigger system}
\label{sec:outrigger}

\begin{figure}
    \centering
    \begin{subfigure}{.61\textwidth}
        \includegraphics[width=\textwidth ]{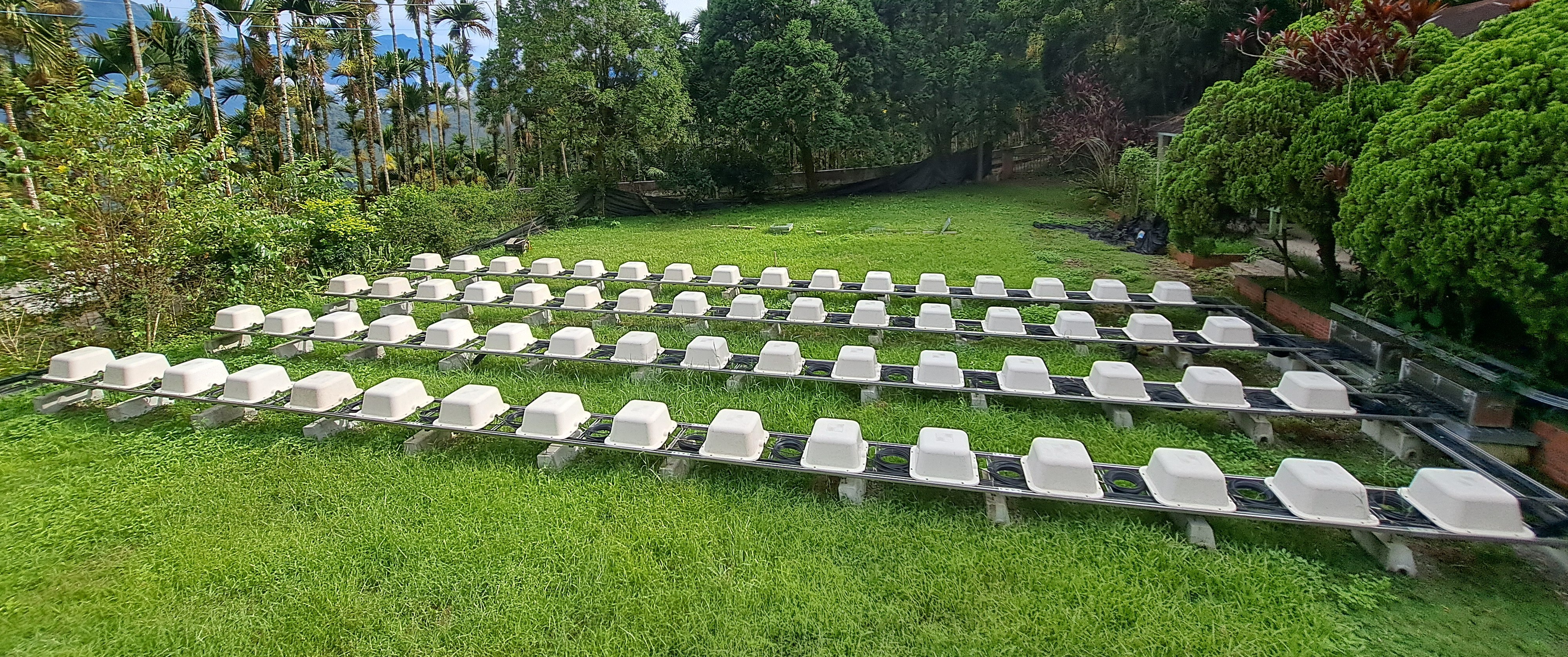}
    \end{subfigure}
     \begin{subfigure}{.32\textwidth}
        \includegraphics[width=\textwidth ]{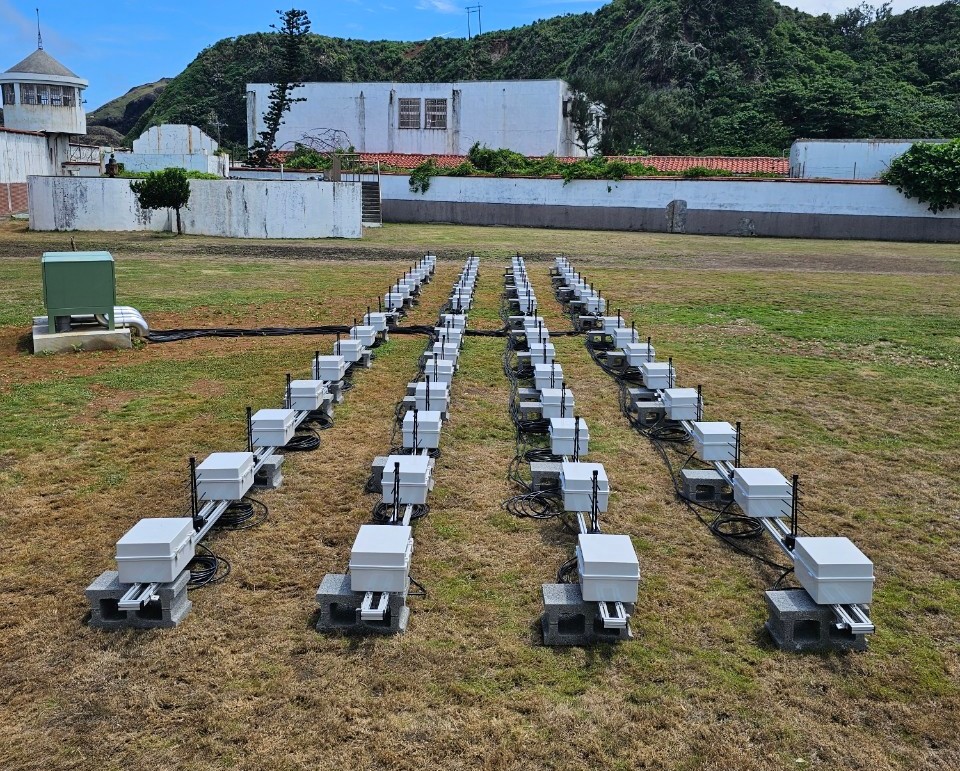}
    \end{subfigure}
    \caption{
       Photos of BURSTT outrigger stations in Taiwan with 64 antennas: LTN at Nantou (left), where the magneto-electric dipole antennas are contained in the white radome; and GRN at Green Island (right).
    }
\label{fig:outrigger}
\end{figure}

Outrigger stations serve the purpose of FRB localization. These stations have a smaller number of antennas than the main station and do not have a pulse-search server. The core function of the outrigger station is the baseband voltage ring buffers. In principle, the outrigger antennas can be deployed into an arbitrary array. Once the baseband data is saved, offline beamforming can still focus the sensitivity to the same direction as that of the triggered beam in the main array. Real-time beamforming is not an absolute requirement.
On the other hand, having a rectangular array and a similar beamforming capability means that one can stream the baseband of a single beam to the hard disk for various VLBI tests. 
As a consequence, the BURSTT outriggers are currently all configured as a $16\times4$ array, with four FPGAs. The beamforming server adopted in the outrigger system is a dual-socket server with Intel Xeon 3rd Gen CPUs and a total of 1~TB of RAM. Data from each FPGA is divided into two UDP packets and stored in two ring buffers on the server. The ring buffer length is set to 16~sec.

%
%
In our initial operation, two outrigger stations are fully operational. The Longtien (LTN) station is located in central Taiwan and uses the magneto-electric dipole (MED) antenna. The LTN station operates in the  \SIrange{300}{700}{\MHz}. A second station in Green Island (GRN), is on a small island off the coast in Eastern Taiwan. The GRN station uses the same LPDA as those used in the main array and also operates in the same \SIrange{400}{800}{\MHz}. 

The data acquisition at outrigger stations also have a fixed \SI{400}{\MHz} bandwidth, with a tunable frequency range between \SI{300}{\MHz} and \SI{800}{\MHz}. This flexibility allows probing FRB signals at lower frequencies, where scattering leads to stronger pulse broadening.

When the pulse-search server at FUS detects a valid pulse signal, it sends out a TCP trigger command with estimated DM and arrival time (in Unix time) to all the outrigger stations to save corresponding baseband data.  

%
%
%
Upon receiving the trigger, outrigger stations will save the full baseband data with FPGA beamforming but without the server beamforming. Since the full baseband is saved, the beamforming direction can be arbitrarily reformed. Thus the triggered baseband can be optimized in offline analysis to obtain a higher SNR \citep{CHIME_baseband_paper}.
The duration of the baseband data for each \SI{200}{\MHz} frequency band is determined by the expected dispersion delay across the band, with an extra margin of $\sim\pm\SI{0.5}{\second}$ to ensure full signal coverage.

\section{Result and Discussion}
\label{sec:result}



The real-time beamforming and pulse search commenced in April 2025 with a partial set of beams and was expanded to the full 256-beam search by late May. The performance of each real-time process was systematically evaluated and verified through the confirmed detections of single pulses of pulsars.

\subsection{Beamforming Verification}

To verify the beamforming described in Sec.~\ref{sec:beamform_FPGA} and \ref{sec:beamform_server}, bright persistent radio sources such as the Sun, Cygnus A, Cassiopeia A, etc., are used as the calibrators.
%
For each target, beamformed intensity data of all beams were recorded for about three hours when the source transited through the beams formed within the field of view, and their temporal variations are compared with the expected ones from simulations.

\begin{figure}
    \centering 
    \includegraphics[width=0.7\textwidth]{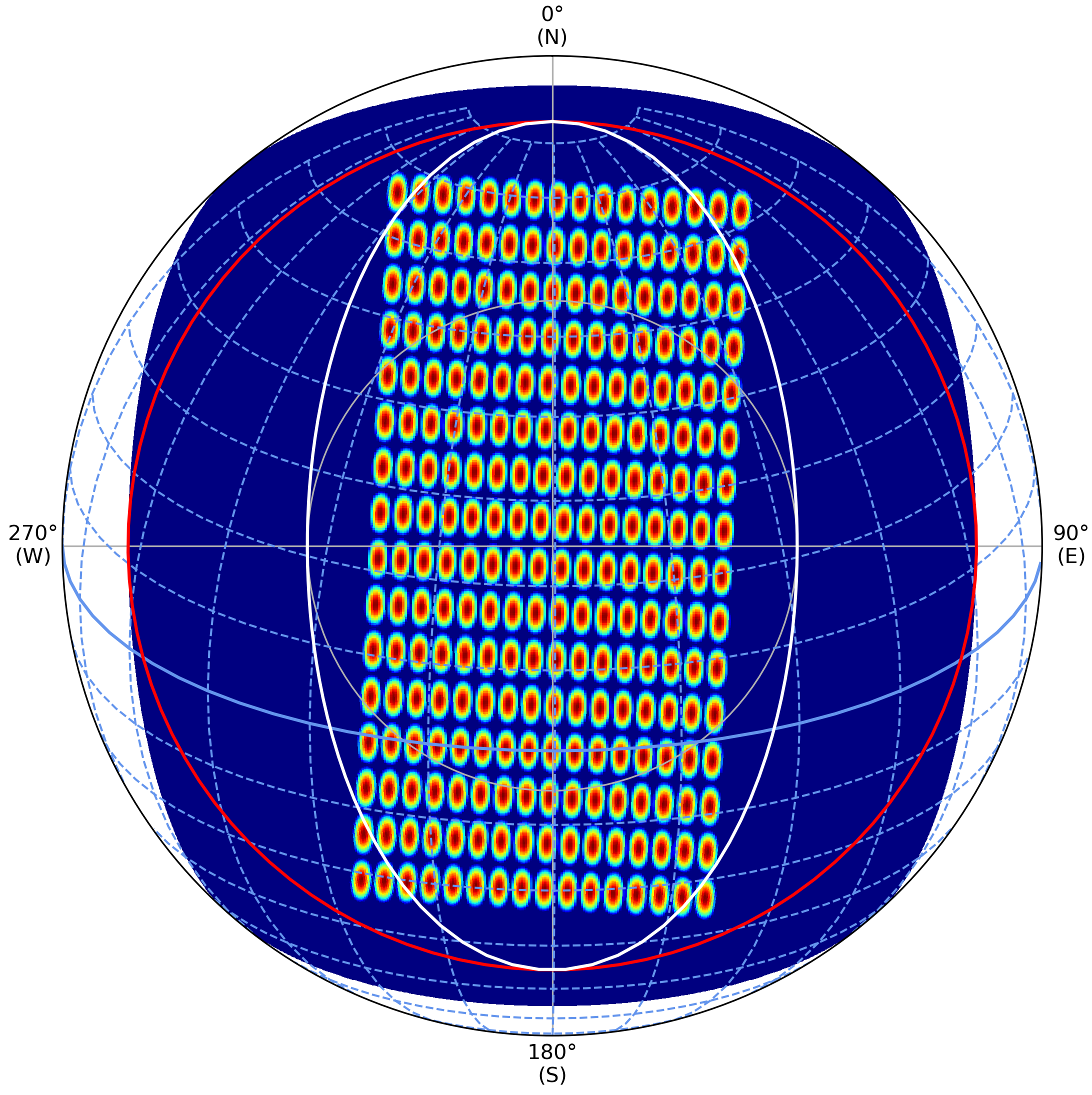}
	\caption{
        The beam distributions with $16 \times 16$ mode. Each beam has a size of $\ang{2.5} \times \ang{5}$. Details on the beamforming matrix is described in \citet{BURSTT_frontend} (in preparation).
	}
\label{fig:256beams}
\end{figure}

For each BURSTT station, beams along East-West (E-W) and North-South (N-S) directions are formed separately (see Sec.~\ref{sec:beamform_FPGA} and \ref{sec:beamform_server}), and they are evenly spaced along each direction by  
\begin{equation}
    \sin\theta_m = \frac{\lambda_0}{d}\frac{m+b_0}{N_{\rm ant}},
    \label{eq:bf_angle}
\end{equation}
where $m = 0, 1,..., (N_{\rm ant}-1)$, and the reference wavelength $\lambda_0$ is chosen to be that of \SI{400}{\MHz}. The antenna spacing $d$ is \SI{1}{\m} and \SI{0.5}{\m} along E-W and N-S directions at the main station, respectively. The  beam offset $b_0$ can be adjusted to maximize the beam overlap between stations for a given target.
The resulting beams at the main station are adjacent to each other, with a beam width of roughly $\ang{2.5} \times \ang{5}$ in E-W and N-S directions, respectively.
The beamforming matrix with 256 beams is described in \citet{BURSTT_frontend} (in preparation) and the beam pattern is shown in Fig.~\ref{fig:256beams}.

\begin{figure}
\centering    \includegraphics[width=1.0\textwidth]{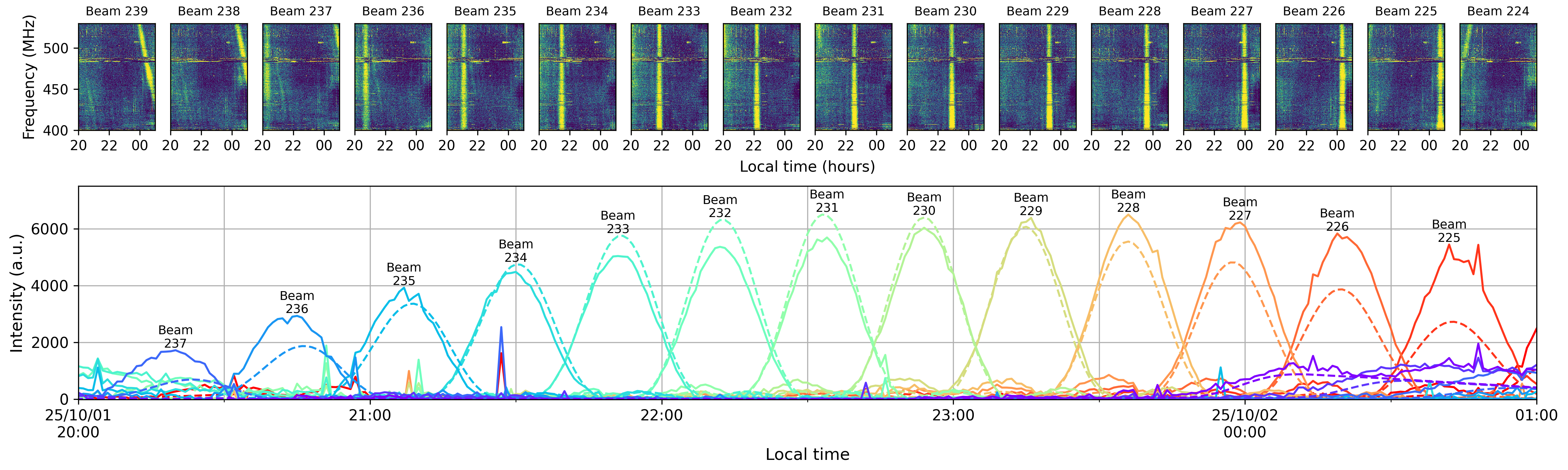}
	\caption{
        The relative beam intensity of Cassiopeia A (RA = 23h23m26s, Dec = +58d48m41s) on Oct 1, 2025. Top panels present the frequency versus time (``waterfall'') plots of the 14th rows (beams 224--239) among $16 \times 16$ beams at the BURSTT main station. Bottom panel shows the frequency-averaged intensity from the beams shown above (solid lines) compared to the simulation (dashed line). The tilted beams seen in beams 224, 238, and 239 are the sidelobes of the Cassiopeia A. 
	}
\label{fig:CasABeam}
\end{figure}

The simulated and observed intensity of Cassiopeia A as a function of time are shown in Fig.~\ref{fig:CasABeam}.
%
Given the known radio flux intensity of Cassiopeia A \citep{Perley2017}, the system equivalent flux density (SEFD) of a single antenna and that after beamforming can be characterized. The details are described in another paper \cite{BURSTT_frontend} (in preparation).

\subsection{Pulse Injection Test}

To verify the performance of the real-time pulse search pipeline under the configuration described in Sec.~\ref{sec:bonsai_conf}, two tests were conducted by injecting FRB-like signals. 
First, disperse pulse signals were generated and injected at the software level using the \texttt{simpulse} package \citep{Merryfield2023}, synthesized together with recorded noise extracted from the intensity data.
%
The simulated FRB signals span a DM range of \SIrange{10}{1000}{\DM} and an SNR range of 0--100. Using these injections, we verified that the pipeline can successfully detect pulses from the up-channelized intensity data and used the results to determine appropriate bonsai parameter settings.

\begin{figure*}
    \centering
    \includegraphics[width=0.32\textwidth]{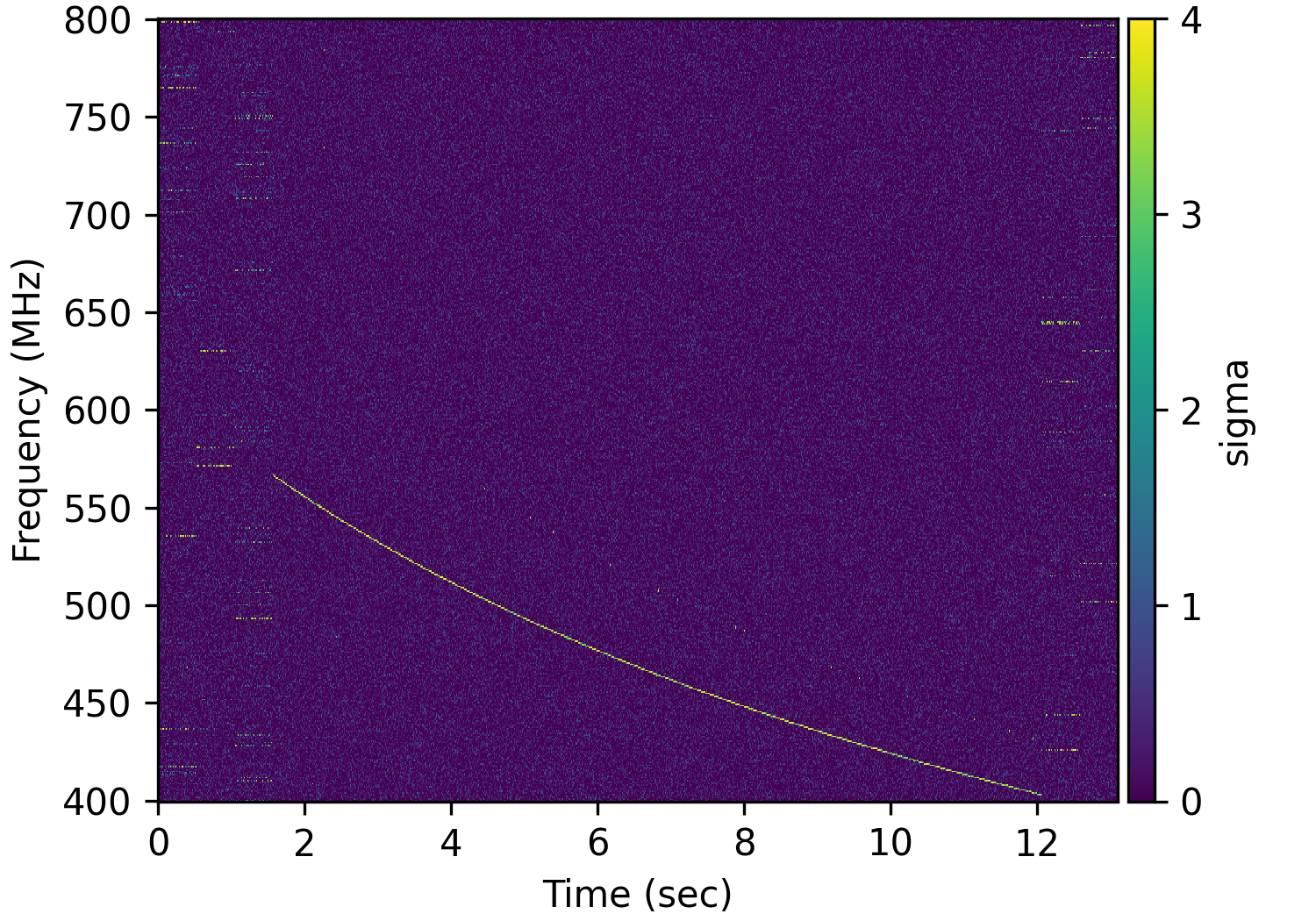}
    \includegraphics[width=0.32\textwidth]{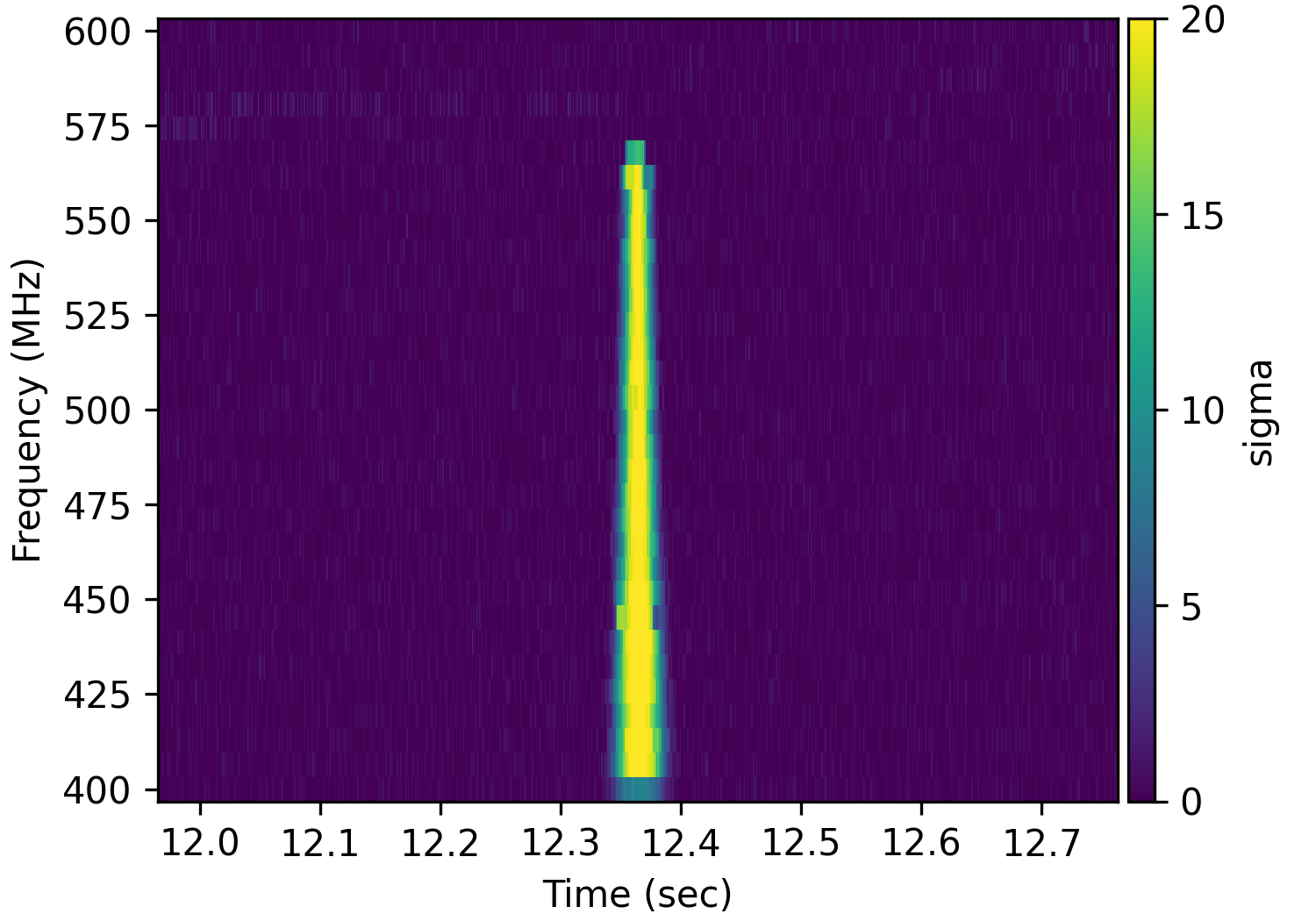}
    \includegraphics[width=0.32\textwidth]{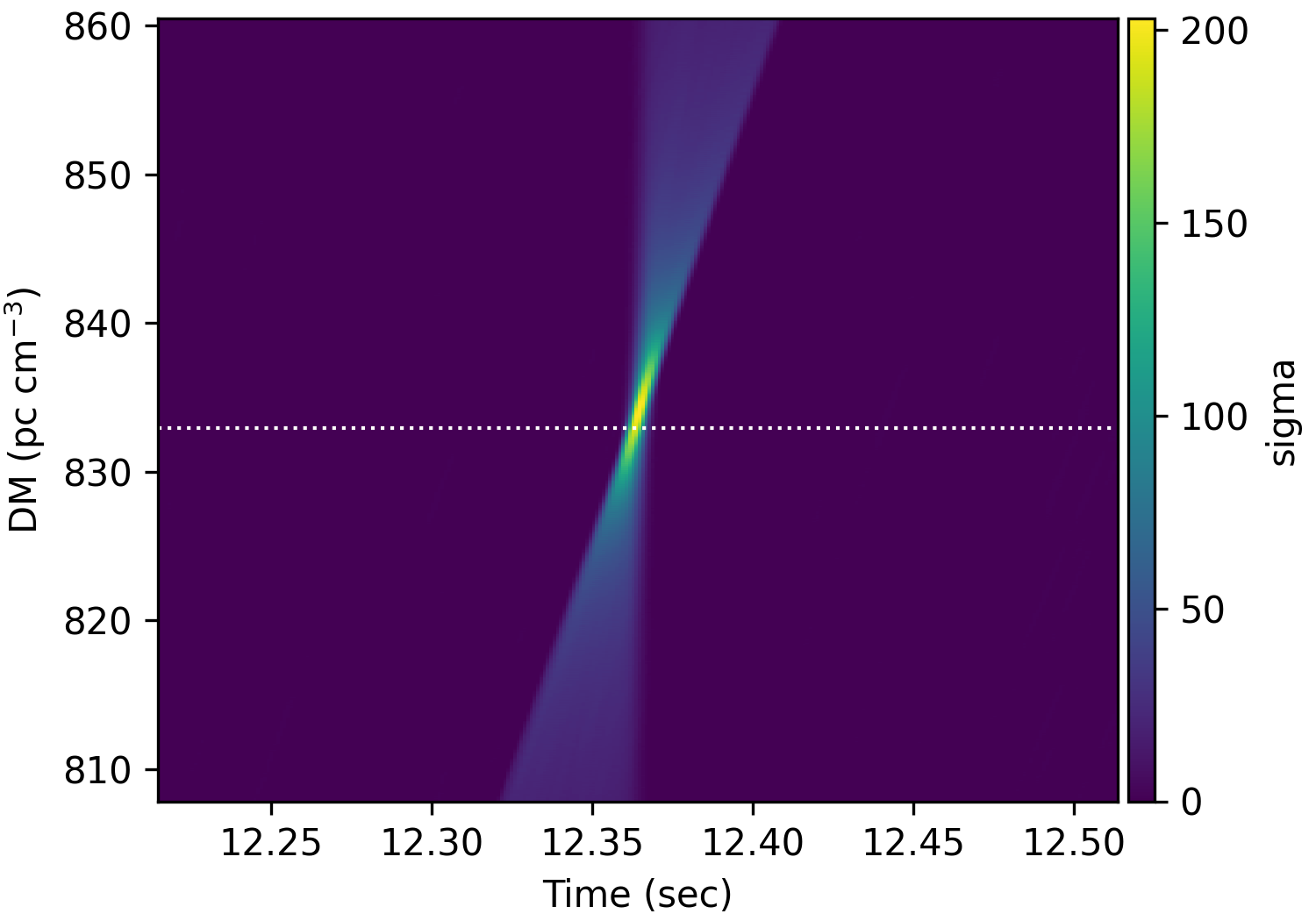}
    \caption{
    The waterfall plot of injected signal with DM $\sim$\SI{800}{\DM} using AWG (left), de-dispersed plot with DM = \SI{833}{\DM} (middle), and DM-time plot from the search pipeline (right).
    }
    \label{fig:AWGInject}
\end{figure*}

The second test utilized an arbitrary waveform generator (AWG) to inject analog dispersed pulses to the FPGA inputs to verify the end-to-end signal path including analog-to-digital conversion, channelization, and beamforming. 
The AWG-generated signal was converted and recorded by the BURSTT backend system. An example of injected pulse event is shown in Fig.~\ref{fig:AWGInject}) demonstrating that the pulses are faithfully reproduced, confirming the high fidelity of the digitization and beamforming processes.

\subsection{Pulse Detection of Known Pulsars}
\label{sec:pulsar}
With 256 antennas currently operational at the main station, the BURSTT backend system regularly detect pulses from two of the brightest pulsars in the northern sky, namely, the Giant Pulses (GPs) from Crab pulsar (PSR B0531+21) and normal pulses from PSR B0329+54 pulsar, identified according to the procedure described in Sec.~\ref{sec:EventID}. 
These sources account for most of the recorded events so far.
Regularly detected pulsar events can not only be used for monitoring the system performance, the pulses detected across stations can be cross-correlated for VLBI calibration.

\subsubsection{Detection of Crab Giant Pulses}
\label{sec:Crab}

The Crab pulsar (RA = 05h34m31.93357s, Dec = +22d00m52.1927s) falls within BURSTT's FoV for approximately three hours per day, with a DM of \SI{56.77}{\DM} \citep{Lyne1993}. One of its most prominent properties is the presence of  narrow, intense giant pulses (GPs) whose fluence can reach a few \si{\kilo\Jy} (e.g., \citealt{Bera2019}). The exceptional brightness and sporadic nature of these GPs make the Crab pulsar an ideal target for evaluating BURSTT’s performance in detecting FRB-like signals. 

\begin{figure}
    \centering
    \includegraphics[width=0.45\textwidth]{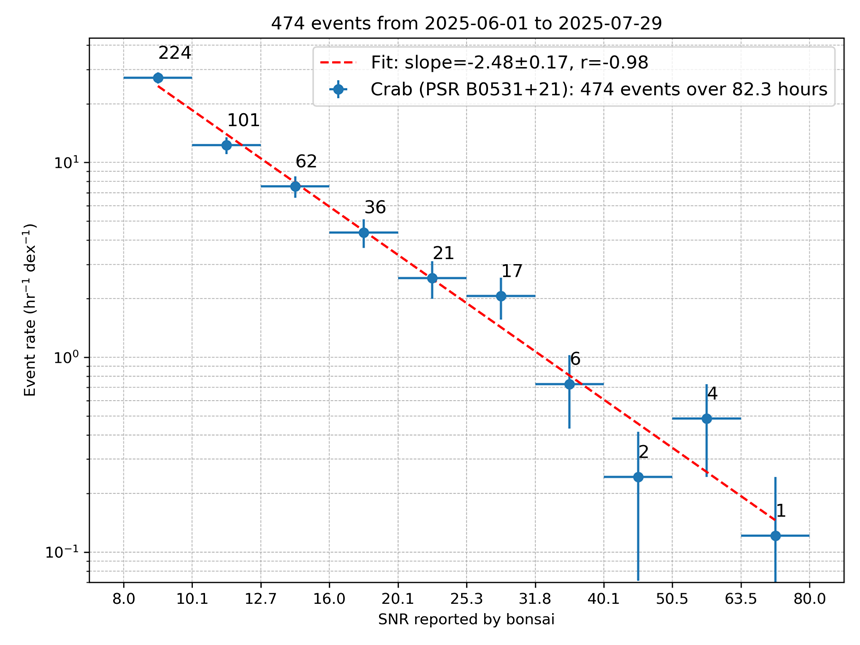}
    \caption{
        The SNR distribution of 474 Crab GPs detected by BURSTT between 2025-06-01 to 2025-07-29. The distribution is consistent with a power law with index of $\alpha=2.48\pm0.17$.
    }
\label{fig:CrabGRPStats}
\end{figure}

As a demonstration, during the period from 2025-06-01 to 2025-07-29, a total on-source exposure of about 178.6 hours was obtained. Candidate events with SNR greater than 8 were retained for further inspection.
%
 Each candidate was manually examined to ensure the detected pulse corresponds to a genuine GP, while sideband detections --- characterized by similar DMs but limited frequency span --- were excluded from the final sample.
In total, 474 Crab GP events were identified in this dataset.
The SNR distribution of the detected Crab GPs during the selected period is shown in Fig.~\ref{fig:CrabGRPStats}. The distribution of event rate ($N$) and SNR ($S$) follows a power-law behavior, with a best-fit power-law index of $\alpha = dlogN/dlogS = -2.48 \pm 0.17$, consistent with previous measurements \citep{Mikami2016,Meyers2017,Bera2019}.


%

\subsubsection{Detecting Normal Pulses from PSR B0329+54}
\label{sec:B0329}
\begin{figure*}
    \centering
    \includegraphics[width=1.0\textwidth]{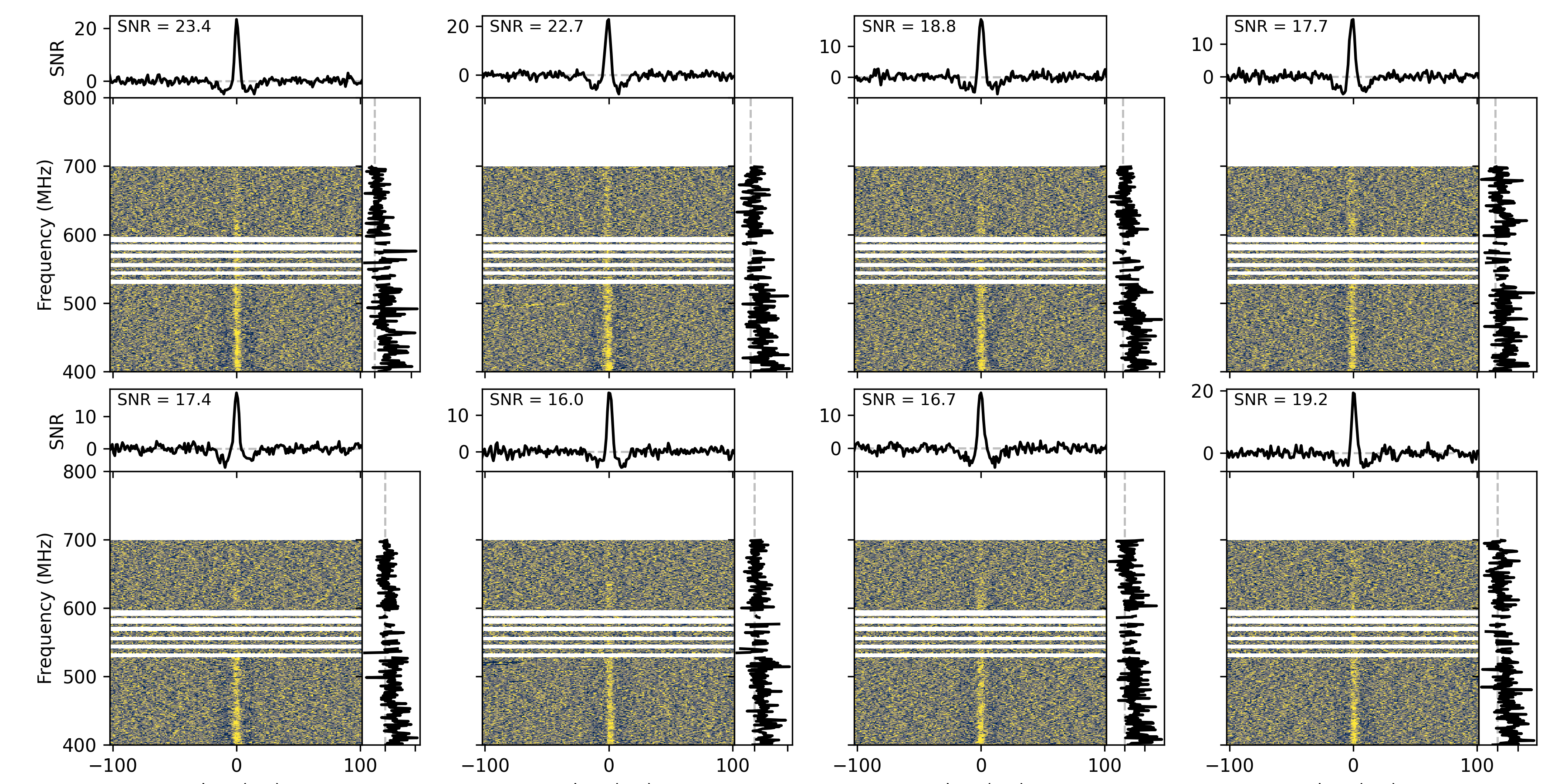}
    \caption{
    Single pulses from PSR B0329+54 detected by the BURSTT real-time pipeline. Similarly as Fig.~\ref{fig:CrabGRP}, each main panel shows a dedispersed dynamic spectrum spanning 400--800 MHz. The frequency-averaged profile (400--700 MHz) is shown above, and the time-averaged spectrum is shown to the right for a 10 ms window around the pulse peak. The intensity data plotted here have a time resolution of $\rm \SI{1.024}{\milli\second}$, with frequency binning of 32 across 16384 channels covering the \SI{400}{\MHz} bandwidth. A rolling mean with a 31~ms window was subtracted from the intensity data. All pulses were detected at $\rm DM \approx \SI{26.7}{\DM}$ and dedispersed at $\rm DM = \SI{26.75}{\DM}$.
    }
    \label{fig:B0329}
\end{figure*}

The B0329+54 pulsar has a DM of \SI{26.76}{\DM} and a rotation period of \SI{0.7145}{\second}  \citep{ATNF_catalogue}.
Its normal pulses have a radio spectrum that peaks at $\sim\SI{300}{\MHz}$ with a flux density of $\sim\SI{3}{\Jy}$  \citep{Kramers2003}.
%
Because of its relatively high flux and short period, the pulsar produces a high rate of detectable pulses. To optimize data handling, only the trigger metadata are recorded without storing the corresponding baseband data, unless the pulsar event has an SNR over 9. 
Examples of detected event after dedispersion are shown in Fig.~\ref{fig:B0329}, where the pulse profiles are clearly visible.

The BURSTT system typically detects about 10 PSR~B0329+54 events each day. The regular detection of this bright pulsar serves both as a benchmark for system sensitivity and stability, and as an opportunity to monitor potential variations in its flux and pulse properties.

\subsection{Receiving Triggers at Outrigger Stations}

\begin{figure}

    \centering
    \includegraphics[width=1.0\textwidth]{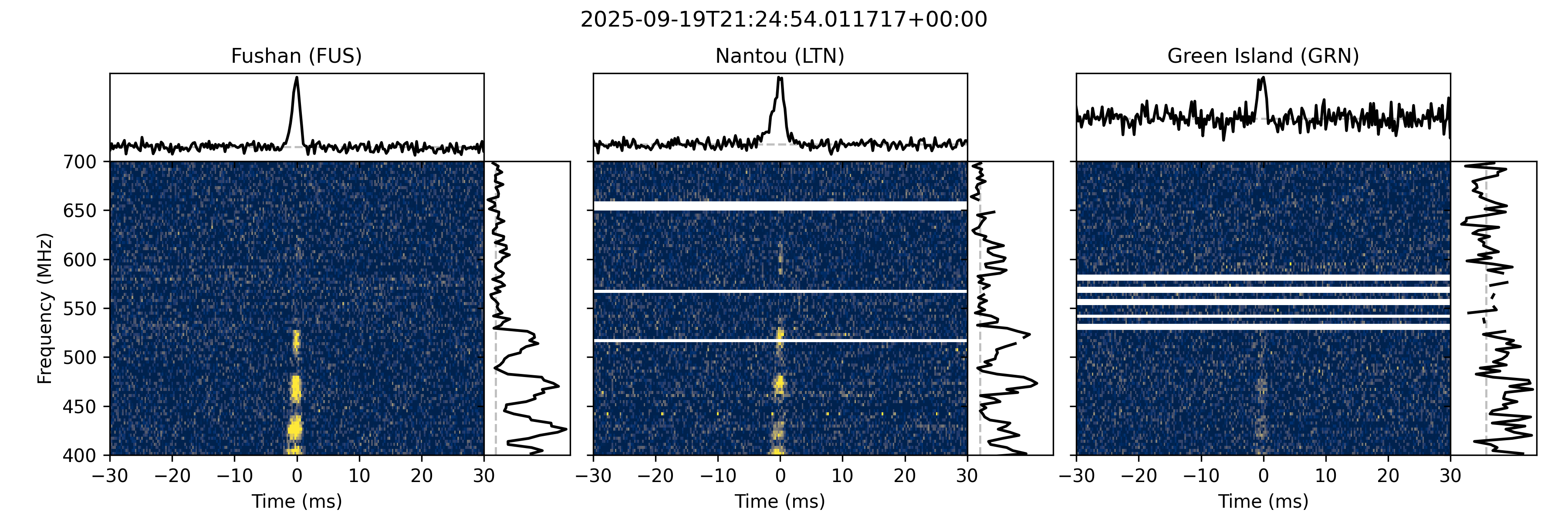}
	\caption{
		An example of Crab Giant Pulse event simultaneously recorded by three stations, the Fushan main station, Nantou and Green Island outrigger stations upon receiving the trigger from real-time search pipeline. The 8 frequency channels are binned among 1024 channels across \SI{8}{\MHz} with the time resolution of \SI{256}{\mathit{\mu}\s}.
	}
\label{fig:CrabGRP2}
\end{figure}

Triggers by Crab GPs are also distributed to BURSTT outrigger stations. An example of simultaneous detection by the main station and two outrigger stations is shown in Fig.~\ref{fig:CrabGRP2}.
Correlations between stations have been measured for many of these simultaneous detections, with early tests suggesting a typical uncertainty of approximately 1 to 2 ns in the delay on the Fushan to Nantou baseline as determined by the fluctuations between nearby pulses.
A critical factor in the VLBI measurement is the relative drift of clocks at different stations. This was addressed by injecting the \SI{10}{MHz} reference signal to a multi-frequency Global Navigation Satellite System (GNSS) receiver. By phase-locking the GNSS receiver to the station clock, the clock error within the precise point positioning (PPP) solution of the GNSS data provides an accurate drift of the station clock against the UTC. We have verified that the PPP clock error between two systems is consistent with that measured directly by a time interval counter (TIC). The uncertainty is about \SI{0.2}{ns} over five hours of monitoring.
However, the system does not currently fully correct for ionospheric effects or the current uncertainty in the exact coordinates of the stations. More detail on the correlations as well as the resolutions to these issues will be discussed in a forthcoming paper on BURSTT VLBI prospects.

\begin{figure}
    \centering
    \includegraphics{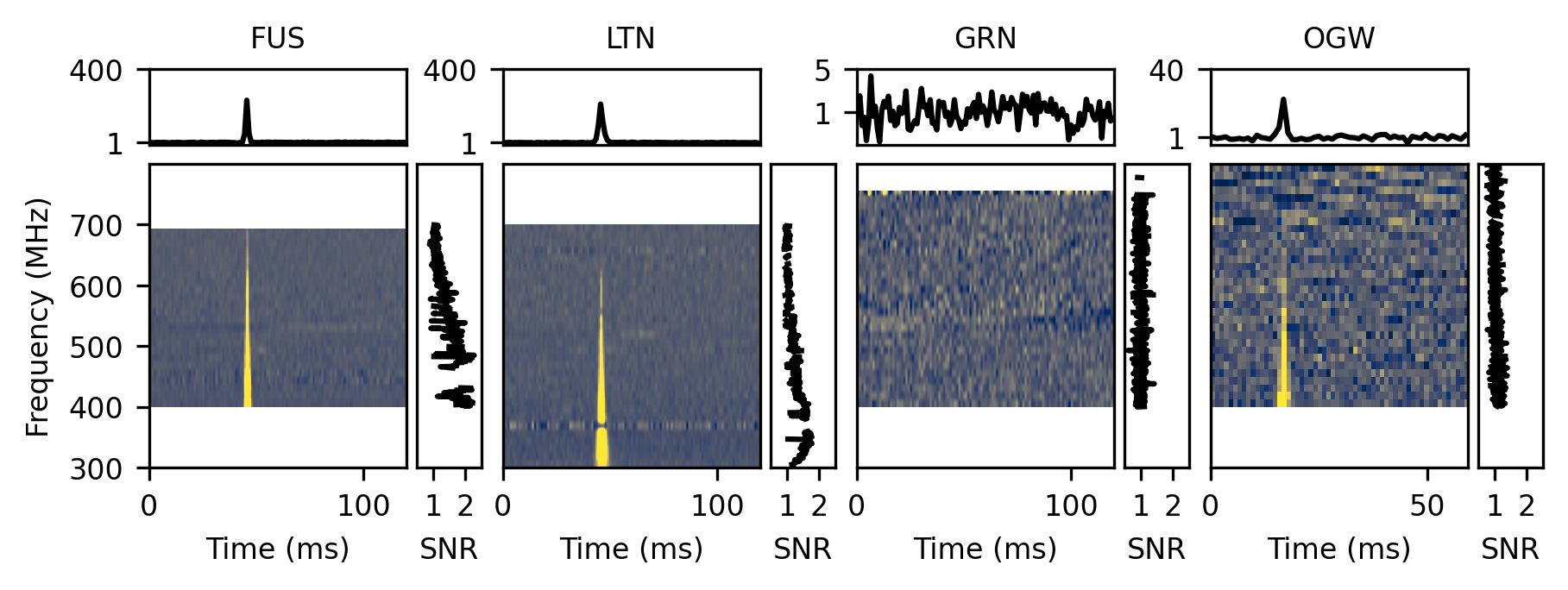}
    \caption{Example Crab giant pulse from 2025-11-28 recorded (from left to right) as Fushan (FUS), Nantou (LTN), Green Island (GRN), and Ogasawara (OGW).}
    \label{fig:20251128_170354_pulses}
\end{figure}

\begin{figure}
    \centering
    \includegraphics{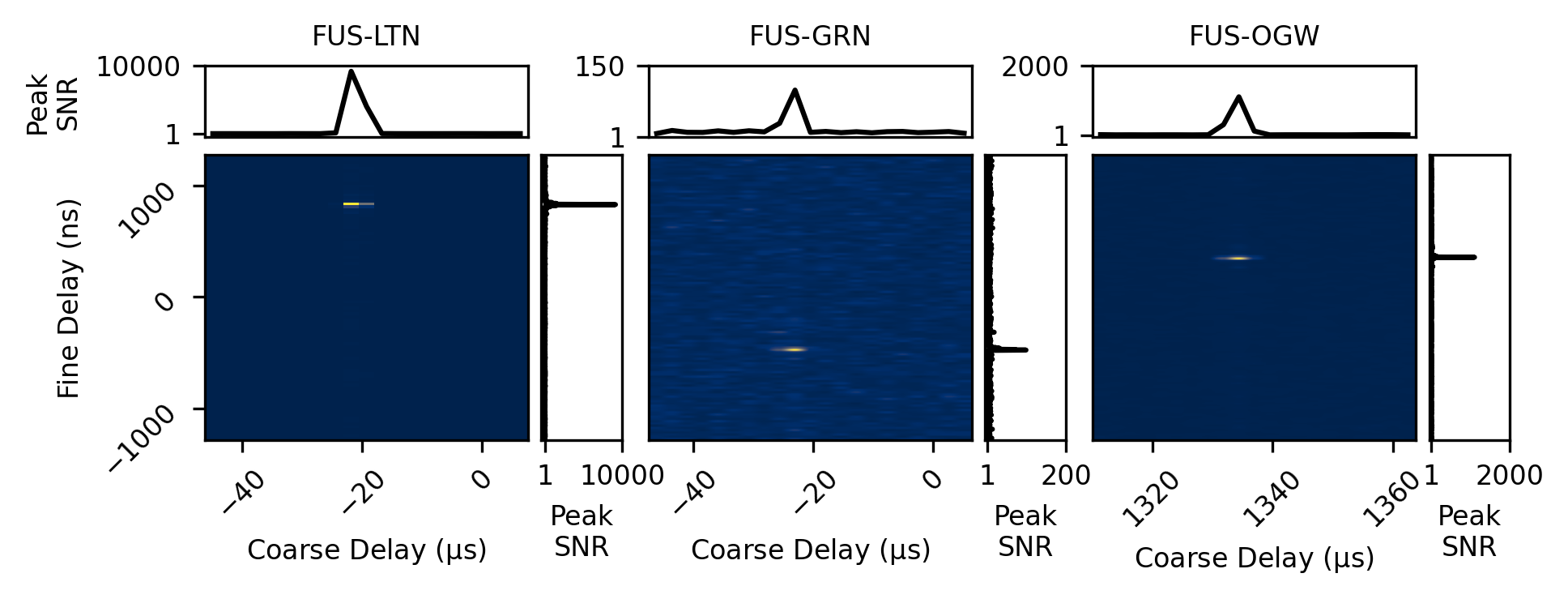}
    \caption{Correlation results between the Fushan station and all three active outriggers for the pulse shown in Fig.~\ref{fig:20251128_170354_pulses} and the peak signal to noise ratio along each axis. The coarse delay axis shows the the amount of time bine shifts in the correlation, while the fine delay shows the Fourier transform along the frequency axis. To improve clarity, the fine delay has been rebinned by a factor of eight using the maximum value.}
    \label{fig:20251128_170354_corr}
\end{figure}

\subsection{False Triggers from RFI}

\begin{figure}
    \centering
    \begin{subfigure}{.31\textwidth}
        \includegraphics[width=\textwidth ]{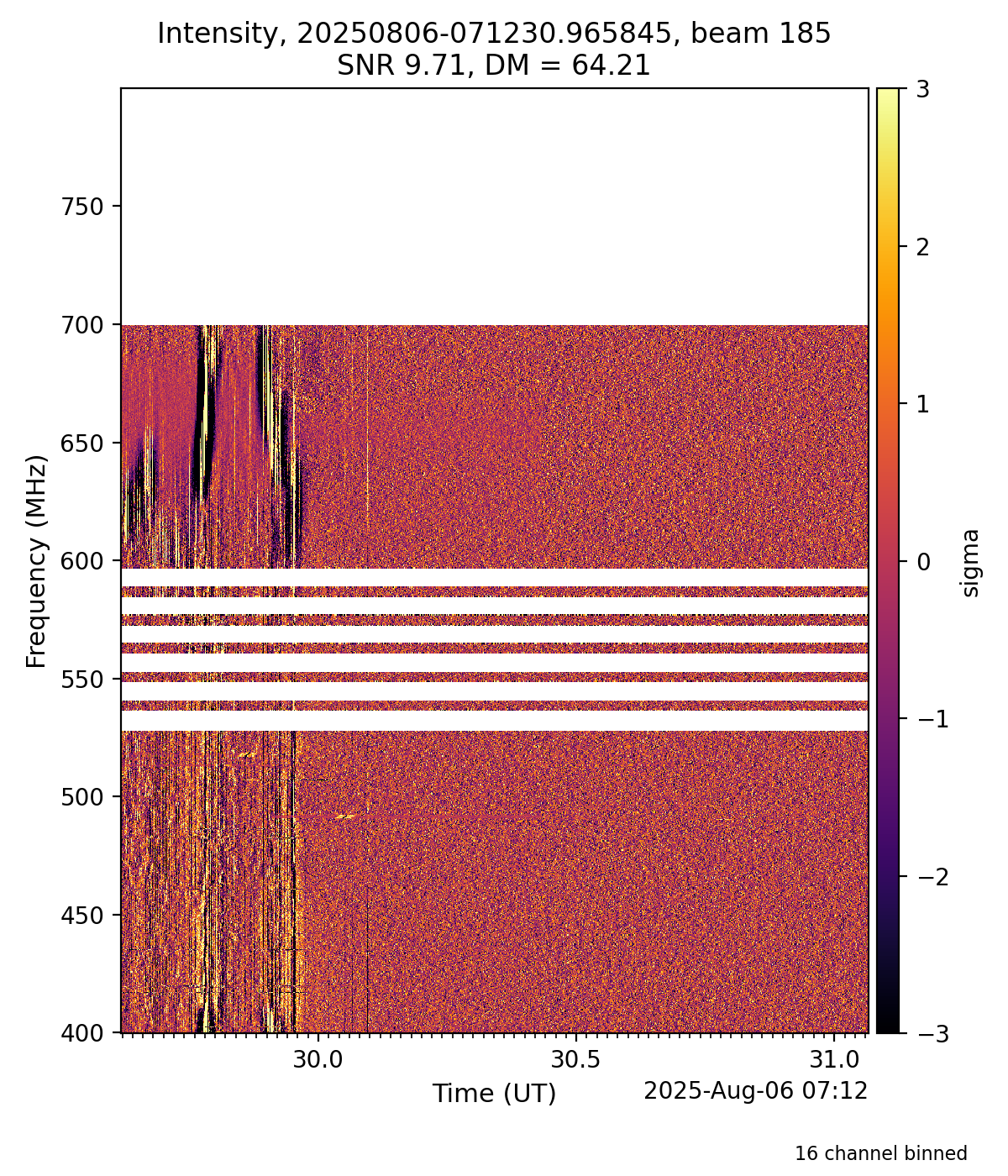}
    \end{subfigure}
     \begin{subfigure}{.31\textwidth}
        \includegraphics[width=\textwidth ]{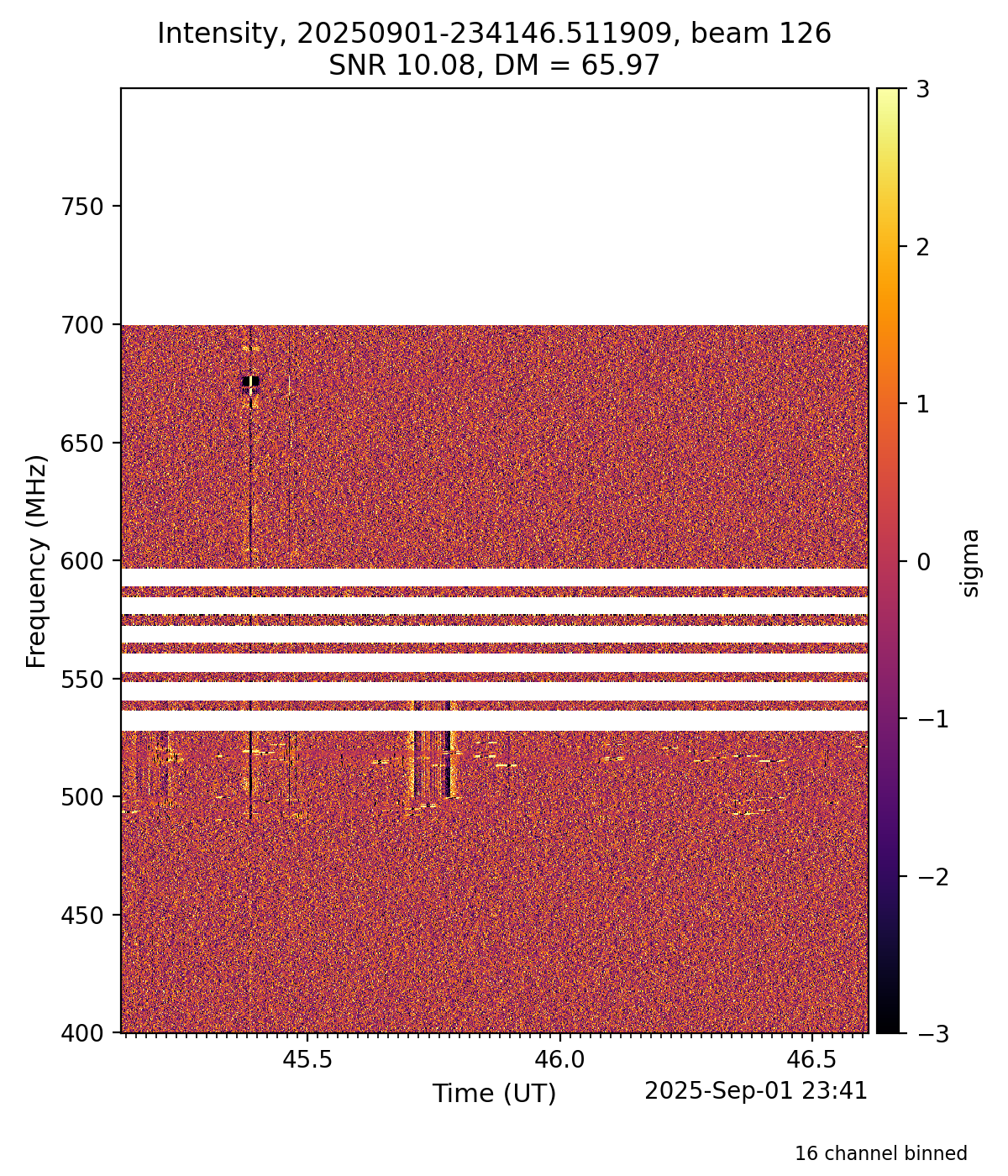}
    \end{subfigure}
    
    \caption{
    Examples of intensity waterfall plots of false-trigger events. Left: an event consisting of multiple pulses within $\sim\SI{300}{\milli\second}$, coincident with local thunderstorm, likely originating from lightning.
    %
    %
    Bottom: a RFI event composed of multiple pulses which cause accidental SNR enhancement at \SI{66}{\DM} where the lower spectral components at $\sim\SI{510}{\MHz}$ get aligned with the higher ones at $\sim\SI{670}{\MHz}$. 
	}
	
\label{fig:RFI}
\end{figure}

Apart from triggers from the pulsars, the BURSTT pulse-search pipeline in its current configuration is also occasionally triggered by various types of RFI. 
%
These false-trigger events typically consist of multiple pulses. During the de-dispersion process of pulse search for optimal DM and arrival time, different spectral components of such signals may, by chance, align in time, resulting in accidental SNR enhancement that exceeds the trigger threshold.
The false-trigger rate is on the order of a few to several events per day, which remains manageable for the current system.
Some false-trigger events were found associated with lightning, satellite, with examples shown in Fig.~\ref{fig:RFI}. As their characteristics are distinct from those of FRB events with a disperse wideband pulse, these events are currently rejected by visual inspection.

%
When thunderstorms occurred within a few tens of kilometers around the main station, the system often detected wideband RFI events passing L1 triggers clustering in time, lasting for a few hours and coincident with the lightning activity recorded by local weather monitor.
It is well known that the lightning emits radio pulses with a time scale from microseconds to milliseconds \citep{Rakov2003}.
Although most of these events were effectively rejected by the trigger requirement of SNR difference relative to the incoherent beam (Sec.~\ref{sec:TrigCriteria}), as lightning acts as an extended source, some events still occasionally passed L2 trigger and caused false triggers.
The detection of such events demonstrates the BURSTT's potential for studying lightning in high temporal (\SI{2.56}{\mathit{\mu}\s}) and spectral (\SI{0.4}{\MHz}) resolution, and for performing lightning mapping if a dedicated pipeline is developed, similar to those accomplished by LOFAR \citep{LOFAR2019}.


There are false-trigger events showing impulsive ($\rm{DM}\approx0$) and narrow-band of $\sim\SI{30}{\MHz}$ bandwidth.
Some of these events only triggered one beam and apparently moved across the sky from east to west in a few hours, suggesting that they might come from satellites.
Such events, when coincident with other pulse at other frequency bands, may produce false triggers (e.g.,~middle panel of Fig.~\ref{fig:RFI}).
%
As ASKAP reported the detection of impulsive events of less than \SI{30}{\ns} duration with an average flux density $>\SI{30}{\kilo\Jy}$ from some decommissioned satellites \citep{ASKAP2025}, a dedicated satellite tracking routine may need to be added to the event identification for flagging such potential RFI. 
As BURSTT's sensitivity will grow with the number of antennas in the near future, a more sophisticated RFI classification is required and being discussed.

\subsection{Throughput and Latency}

We summarize the end-to-end throughput and latency of the BURSTT backend system in Table~\ref{tab:latency}. The total latency, defined as the time from signal arrival at the FPGAs to generation of an FRB candidate by the dedispersion pipeline, is typically less than $\sim$3.5~s. This interval encompasses primary and secondary beamforming, data transfer to the server cluster, time integration at 1.024~ms cadence, real-time pulse search using \texttt{bonsai}, and trigger generation. The pipeline operates continuously at the full input data rate without packet loss, demonstrating that the system can sustain real-time processing for all 256 beams and the full DM search range. Additionally, Table~\ref{tab:power} summarizes the power consumption of the back-end system.

\begin{deluxetable}{lccc}
\renewcommand{\arraystretch}{1.35}
\tablecaption{End-to-end throuput and latency of the BURSTT backend system \label{tab:latency}}
\tablehead{
\colhead{Stage} & 
\colhead{Throughput per node} & 
\colhead{Output} & 
\colhead{Latency}
}
\startdata
\hline
FPGA channelization & 51.2 GB/s & Voltage streams & -- \\
Primary beamforming (FPGA) & 51.2 GB/s (per FPGA) & 16 beams (per FPGA) & $<2.5$~ms \\
Data transfer to servers & 6.45 GB/s (per server) & Voltage/intensity streams & $\sim$ a few ms \\
\hline\\[-3.5mm]
\makecell[l]{Secondary beamforming (server)\tablenotemark{a} \\ ~~~\& integration} & 25.8 GB/s & \makecell[c]{Intensity streams of 256 beams \\ (1.024 ms $\times$ \SI{390.625}{\kHz})} & $\sim$ 10 ms \\ 
Preprocessing & 1.0 GB/s & Intensity streams (1.024 ms $\times$ \SI{24.414}{\kHz}) & $\sim$ 0.3 s \\
\hline
Intensity stream receiving\tablenotemark{b} &  4.0 GB/s & -- & $\sim$ 0.1 s \\
Pulse search (\texttt{bonsai}) & 12 MB/s (per beam) & Candidate list per beam & $\leq$ 1.6 s \\
Multi-beam trigger selection & -- & FRB alert & $\sim$ 1.5 s \\
\hline
Total end-to-end & & FRB candidate & $\sim$ 3.5 s \\
\enddata
\tablenotetext{a}{Each of the 4 beamforming servers handles the equivalent processing load of 4 FPGAs.}
\tablenotetext{b}{The \texttt{bonsai} server receives intensity streams from all four beamform servers.}
\end{deluxetable}

\begin{deluxetable}{lccc}
\renewcommand{\arraystretch}{1.35}
\tablecaption{Power Consumption of the BURSTT system \label{tab:power}}
\tablehead{
\colhead{Node} & 
\colhead{Per unit (W)} & 
\colhead{Number} & 
\colhead{Total (kW)}
}
\startdata
Front-end & 5 & 256 & 1.3 \\
FPGA & 120 & 16 & 1.9 \\
Beamform servers & 1100 & 4 & 4.4 \\
Pulse search server & 1100 & 1 & 1.1\\
Network switches & 200 & 2 & 0.4 \\
\hline
Sum & & & 9.1\\
\enddata
\end{deluxetable}

\subsection{Other Design Considerations}\label{sec:lessons}

To scale up the BURSTT-like back-end system, we consider that the computation required for the factorized beamform is $N_{1st_beam} \times N_{2nd_beam} \times N_{FPGAs}$. For a larger array, the number of FPGAs needed scales linearly with $N_{ant}$. If we again set $N_{beam} = N_{1st_beam}\times N_{2nd_beam}$ to be of the order of $N_{ant}$, then the computation in CPUs scales quadratically with $N_{ant}$. More CPU nodes will be needed to share the computation, with each node handling a smaller chunk of the bandwidth. On the other hand, the current design has not reached the throughput limit of the FPGAs and CPUs. The team is continually developing the back-end software to fully utilize the computation capacity.

Another important consideration is that the sensitivity varies a lot, by $\lambda^2$, from \SIrange[]{400}{800}{MHz}. This means the array spacing optimized for the lower frequency would be too thinly populated at the higher frequency, or vice versa. Effectively, only about \SI{200}{MHz} of bandwidth is really useful. Therefore, a more efficient design is to frequency-convert and combine multiple dipoles into one ADC channel of the FPGA, with a smaller bandwidth per antenna. The FFT channelization resource in the FPGA is unchanged. However, it would require a larger matrix multiplication in the FPGA and usage of the full bandwidth of the 100~GBE output interface. This is an ongoing work and will be reported in the project's next phase.

\section{Conclusion}

System performance has been validated through beamforming tests using bright radio sources such as the Sun and Cassiopeia A, and through both software-based and AWG-injected pulse tests confirming the high fidelity of the signal-processing pipeline. BURSTT has successfully achieved real-time detections of bright giant pulses from the Crab pulsar and regular pulses from PSR B0329+54, typically detecting about ten events per day from the latter, thus providing robust benchmarks for system sensitivity and stability.

With its wide field of view ($\sim60^{\circ} \times 120^{\circ}$) and sub-arcsecond localization capability, BURSTT is optimized for detecting fast radio bursts (FRBs) in the nearby Universe (median redshift $\approx 0.04$), with an expected rate of roughly 50 detections per year---offering valuable opportunities to study the properties and origins of FRBs in unprecedented detail. Future developments will integrate multiple outrigger stations across baselines exceeding 100 km for joint interferometric localization, and the operational band will be optimized to 300–700 MHz to minimize spectral interference. In addition, BURSTT’s high temporal and spectral resolution makes it highly sensitive to solar radio bursts and transient terrestrial radio phenomena such as lightning, extending its potential applications to solar, atmospheric, and RFI-related studies.



\section*{Acknowledgement}
BURSTT is support by Academia Sinica by AS-GCS-114-M02 and by Ministry of Science and Technology of Taiwan by NSTC 113-2923-M-001-003-MY3.

\bibliographystyle{aasjournalv7}
\bibliography{reference}

\end{document}